\documentclass[prd,notitlepage,showpacs,preprintnumbers,amsmath,amssymb,nofootinbib,aps,10pt,superscriptaddress,twocolumn]{revtex4-2}

\usepackage{dcolumn}
\usepackage{bm}
\usepackage{xcolor}
\usepackage[utf8]{inputenc}
\usepackage[spanish,english]{babel}
\usepackage{amsmath,amssymb,amsfonts,latexsym,cancel}
\usepackage{graphicx}
\usepackage{color}
\usepackage{soul}
\usepackage{ulem}
\usepackage{amsmath}
\usepackage{slashed}
\usepackage{braket}
\usepackage{amssymb}
\usepackage{amsmath}
\usepackage{enumitem}
\usepackage{hyperref}
\allowdisplaybreaks % allows to break align equations in different pages. If we want this to not happen between two given lines, use "\\*" to break lines instead of "\\".


\newcommand{\be}{\begin{equation}}
\newcommand{\ee}{\end{equation}}
 \newcommand{\bea}{\begin{eqnarray}}
\newcommand{\eea}{\end{eqnarray}}

\newcommand{\mpl}{m_{\rm pl}}
\newcommand{\mpls}{m_{\rm pl}^2}
\newcommand{\CL}{{\tt ${\mathcal C}$osmo${\mathcal L}$attice}~}

\begin{document}

\title{Equation of state during (p)reheating with trilinear interactions}

\newcommand{\addressUNIBAS}{Department of Physics, University of Basel, Klingelbergstr. 82, CH-4056 Basel, Switzerland.}

\newcommand{\addressIFIC}{Instituto de Fisica Corpuscular (IFIC), Universitat de Valencia (UV) and\\
\it Consejo Superior de Investigaciones Cientificas (CSIC), 46980, Valencia, Spain}

\newcommand{\addressICCUB}{Departament de Física Quàntica i Astrofísıca \& Institut de Ciències del Cosmos (ICCUB),
Universitat de Barcelona, Martí i Franquès 1, 08028 Barcelona, Spain}

\author{Stefan Antusch}
\affiliation{\addressUNIBAS} 
\author{Kenneth Marschall}
\affiliation{\addressIFIC}
\author{Francisco Torrenti}
\affiliation{\addressICCUB}

\date{\today}

\begin{abstract}
We characterize the post-inflationary evolution of the equation of state of the universe from the end of inflation until the onset of radiation domination, when the inflaton is coupled to a daughter field through a trilinear interaction. We consider an inflaton potential that is quadratic near the minimum and flattens in the inflationary regime. By simulating the dynamics in 2+1-dimensional lattices, we have tracked the long-term evolution of the equation of state for about ten e-folds of expansion, for various coupling strengths. The trilinear interaction initially excites daughter field modes through a process of tachyonic resonance immediately after inflation and triggers a temporary deviation of the equation of state from $\bar{w} =0$ to a maximum value $\bar{w} = \bar{w}_{\rm max} < 1/3$. However, at much later times, the inflaton homogeneous mode once again dominates the energy density, pushing the equation of state towards $\bar{w} = 0$ until the onset of perturbative reheating. By combining the lattice results with a Boltzmann approach, we characterize the entire post-inflationary expansion history, which allows to calculate precise predictions for the inflationary CMB observables. We also accurately compute the redshift of the stochastic gravitational wave background produced during preheating, and show that taking the temporary return of the equation of state towards $\bar{w} = 0$ into account can reduce the amplitude by many orders of magnitude relative to previous estimates.
\end{abstract}

\maketitle

\section{Introduction}

Compelling evidence \cite{Planck:2018jri} supports the existence of an early stage of accelerated expansion of the universe called inflation \cite{Starobinsky:1980te,Guth:1980zm,Linde:1981mu,Albrecht:1982wi}. In its simplest realization, inflation is driven by a slow-rolling scalar field called the inflaton, with a potential capable of sustaining the accelerated expansion for a sufficiently long time. Although a plethora of inflationary models have been developed over the past few decades based on different high energy physics models, precision measurements of the CMB anisotropies \cite{Planck:2018jri,BICEP:2021xfz,ACT:2025fju} provide an excellent opportunity to constrain the range of observationally viable inflationary scenarios. In particular, given a specific potential, one can compute the \textit{scalar tilt} $n_s$, which encodes the mild scale dependence of the primordial scalar power spectrum, as well as the tensor-to-scalar ratio $r$, encoding the amplitude of the so-far undiscovered primordial gravitational wave background. The value of $n_s$ has been constrained with an accuracy of less than one percent, while an upper bound has been determined for $r$ \cite{BICEP:2021xfz}.

Inflationary predictions for CMB observables have a theoretical uncertainty stemming from the unknown expansion history between the end of inflation and the onset of a radiation-dominated universe, parametrized by the equation of state \cite{Dai:2014jja,Martin:2014nya,Munoz:2014eqa,Gong:2015qha,Cook:2015vqa}. This uncertainty may typically exceed the experimental errors. To precisely determine its post-inflationary evolution, one needs to fully characterize reheating, which, in its most complete form, must guarantee a complete transfer of energy from the inflaton to a thermal distribution of Standard Model fields, before Big Bang Nucleosynthesis begins at $T\:{\sim}\:1\;{\rm MeV}$ \cite{Kawasaki:1999na,Kawasaki:2000en,Hannestad:2004px,Hasegawa:2019jsa}. A complete characterization of reheating is challenging, as the details depend strongly on the assumed high-energy physics. However, general features frequently emerge. The early stages of reheating are typically dominated by an out-of-equilibrium excitation of field fluctuations dubbed \textit{preheating} \cite{Traschen:1990sw,Kofman:1994rk,Shtanov:1994ce,Kaiser:1995fb,Kofman:1997yn,Greene:1997fu,Kaiser:1997mp,Kaiser:1997hg} (see \cite{Allahverdi:2010xz,Amin:2014eta,Lozanov:2019jxc,Allahverdi:2020bys} for reviews), which leads to the fragmentation of the inflaton. This is followed by a stage of turbulence \cite{Micha:2002ey,Micha:2004bv}, after which a thermalized distribution of Standard Model species must form. Non-linearities become highly relevant once the inflaton fragments, so in order to accurately resolve the post-inflationary dynamics, one typically needs to resort to fully-fledged lattice field simulations \cite{Figueroa:2020rrl,Barman:2025lvk}. Lattice techniques have been commonly employed to study the post-inflationary inflaton fragmentation and equation of state, see e.g.~\cite{Podolsky:2005bw,Dufaux:2006ee,Figueroa:2016wxr,Maity:2018qhi,Saha:2020bis,Garcia:2023eol,Garcia:2023dyf,Saha:2024lil} for the case of monomial inflaton potential with interactions to daughter fields, as well as \cite{Child:2013ria,Krajewski:2018moi,Nguyen:2019kbm,vandeVis:2020qcp,DeCross:2015uza, DeCross:2016fdz, DeCross:2016cbs} for similar studies in related models. Remarkably, preheating also produces a stochastic background of gravitational waves (GWs) \cite{Easther:2006gt,GarciaBellido:2007dg,GarciaBellido:2007af,Dufaux:2007pt,Dufaux:2008dn,Dufaux:2010cf,Bethke:2013aba,Bethke:2013vca,Figueroa:2017vfa,Adshead:2018doq,Adshead:2019lbr,Adshead:2019igv}, but in order to accurately predict the GW background as it would be measured today, one needs to redshift it from the early universe until now, which also requires a complete understanding of the post-inflationary expansion history. 

In this work we focus on inflaton potentials that are quadratic around a minimum. In general, inflaton potentials that behave as $V(\phi) \propto |\phi|^p$ with $p \hspace{-0.06cm} \geq \hspace{-0.06cm} 2$ are compatible with the observational upper bound of the tensor-to-scalar ratio, as long as they flatten at sufficiently large amplitudes. Immediately after inflation, the inflaton is almost completely homogeneous, and its oscillations around the minimum generate the effective (i.e.~oscillation-averaged) equation of state $\bar{w}_{\rm hom} \equiv (p-2)/(p+2)$ \cite{Turner:1983he}. For $p \hspace{-0.06cm} > \hspace{-0.06cm} 2$, the inflaton unavoidably fragments due to self-resonant effects, triggering a transition towards radiation-domination $\lesssim \hspace{-0.06cm} 10$ e-folds after the end of inflation \cite{Lozanov:2016hid,Lozanov:2017hjm}.\footnote{Note that for $p>2$, the existence of sufficiently strong interactions between the inflaton and other fields may significantly speed up the fragmentation process and the achievement of radiation domination \cite{Antusch:2020iyq,Antusch:2021aiw}, so this number must be taken as an upper bound.} For $p=2$, the inflaton survives instead as an oscillating homogeneous condensate.\footnote{Note that for $p=2$, the inflaton may fragment due to gravitational effects even in the absence of interactions with other fields \cite{Musoke:2019ima}, but at much later times than in the case of preheating.} 

The existence of inflaton interactions to other fields, which are expected due to the need to reheat the universe, can fundamentally change this picture. For example, one could consider an inflaton coupled to an (effectively massless) daughter field $X$ through a \textit{quadratic-quadratic} interaction $\phi^2 X^2$. If the coupling is sufficiently strong, the latter gets excited through a process of parametric resonance \cite{Kofman:1994rk,Kofman:1997yn,Greene:1997fu}, triggering the fragmentation of the inflaton condensate through backreaction effects a few e-folds after the end of inflation. Alternatively, or additionally, one could also consider a \textit{trilinear} interaction $\phi X^2$. In this case, the effective mass of the daughter field becomes negative during half of each oscillation, which excites the daughter field through a process of tachyonic resonance, and can trigger an even faster inflaton fragmentation \cite{Dufaux:2006ee}. Furthermore, with only quadratic-quadratic interactions of the type $\phi^2 X^2$, some energy always remains in the inflaton even at very late times \cite{Kofman:1997yn,Figueroa:2016wxr,Antusch:2020iyq,Antusch:2021aiw,Antusch:2022mqv}, i.e.\ reheating remains incomplete. Adding trilinear interactions $\phi X^2$ resolves this shortcoming by introducing a perturbative decay channel that allows reheating to complete by fully transferring the inflaton energy to the daughter fields. This raises various questions, for instance: How, and under which conditions, does the inclusion of trilinear interactions modify the evolution of the equation of state during the preheating phase, with/without various other interactions being present? How does the strength of the effects from the trilinear interaction relate to the time when the perturbative decay happens? And finally: How can we obtain a complete evolution of the equation of state that includes lattice results for the preheating stage as well as the perturbative decay? In this work we address these questions. 

Preheating with trilinear interactions was originally studied with lattice simulations in \cite{Dufaux:2006ee}, which observed a stabilization of the equation of state around $\bar{w} \sim 0.25$ after inflaton fragmentation. However, their simulations were not long enough to observe the later evolution. In this work, we surpass this circumstance by simulating the post-inflationary dynamics in 2+1-dimensional lattices, which allow us to capture up to $\lesssim 10$ e-folds of expansion. This technique was used in previous works for the case of $\phi^2 X^2$ interactions \cite{Antusch:2020iyq,Antusch:2021aiw,Antusch:2022mqv}, and allows to realistically mimic the real three-dimensional dynamics at a lower computational cost (this is demonstrated in App.~\ref{App:3D} of this work). Anticipating slightly our results, these simulations will allow us to observe the relaxation of the equation of state back towards $\bar{w} \rightarrow 0$ after preheating ends. The later evolution will be resolved by solving an appropriate set of Boltzmann equations, which will enable us to completely characterize the evolution of the equation of state until the completion of perturbative reheating and the definitive transition towards a radiation-dominated universe. Here, we also leverage our results to significantly improve the accuracy of two observational predictions of the model considered. First, we provide accurate predictions for the inflationary CMB observables for various coupling strengths. Second, we compute the redshift of the stochastic gravitational wave background produced during preheating \cite{Cosme:2022htl} and update predictions for the frequency and amplitude of the main spectral peak as would be measured today.

The structure of this work is as follows. In Sect.~\ref{Sec:Model} we describe the details of the considered model. In Sect.~\ref{sec:LinearAn} we study the initial preheating stage with trilinear interactions within the scope of a semi-analytical linear analysis. In Sect.~\ref{Sec:Lattice} we present our results from 2+1-dimensional lattice simulations, which fully capture the non-linearities of the system. In Sect.~\ref{sec:RehCMB} we study the later stage of perturbative reheating by solving the corresponding Boltzmann equations from the end of the simulation until the achievement of radiation-domination. By combining the results of Sect.~\ref{Sec:Lattice} and \ref{sec:RehCMB}, we obtain a full characterization of the post-inflationary expansion history. The following two sections study two observational implications of our results: in Sect.~\ref{sec:CMBconst} we obtain predictions for the inflationary CMB observables, and in Sect.~\ref{sec:GWsignatures} we provide accurate predictions for the GW spectrum from preheating measured today. In Sect.~\ref{sec:summary} we discuss our results and conclude. In App.~\ref{App:Quadratic} we consider a model in which the inflaton is coupled to a daughter field through both trilinear and quadratic-quadratic interactions, expanding this way the scenario considered in the main text. In App.~\ref{App:Technical} we provide further technical details.

\section{Considered model}\label{Sec:Model}

In this work we study the post-inflationary dynamics of a model consisting of an inflaton field $\phi$ and a massless scalar \textit{daughter} field $X$, coupled to the inflaton through a trilinear (three-leg) interaction $\propto\!\phi X^2$. We consider the following potential,
\be
V (\phi,X) = V_{\rm inf} (\phi) + \frac{1}{2}h\phi X^2 + \frac{1}{4}\lambda X^4 \ , \label{eq:model}
\ee
where $V_{\rm inf} (\phi)$ is the inflaton potential, $h$ is the coupling constant of the trilinear interaction with dimensions of mass, and we have added a quartic self-interaction to the daughter field for stability purposes, with $\lambda$ being dimensionless. The trilinear interaction serves as a portal to transfer energy from the inflaton to the daughter field after inflation. This may happen either \textit{i)} through non-perturbative processes that emerge during the early reheating stage, i.e.~the phase of \textit{preheating}, or \textit{ii)} via the perturbative decay $\phi \rightarrow XX$, which leads to the reheating of the universe.

In this work we consider inflaton potentials that are quadratic around the minimum, 
\be
V_{\rm inf} (\phi)|_{\min}\approx V_{\rm m} (\phi) \equiv \frac{1}{2} m_{\phi}^2 {\phi}^2 \ .  \label{eq:mon-pot} \\
\ee
The current upper bound of the tensor-to-scalar ratio of the CMB perturbations rules out the possibility of the monomial function \eqref{eq:mon-pot} describing the inflaton potential at all field amplitudes \cite{Planck:2018jri}. However, one can envisage a potential that is quadratic around the minimum, but flattens at large amplitudes to remain compatible with cosmological observations. One possibility is the symmetric $\alpha$-attractor T-model \cite{Kallosh:2013hoa},
\be 
V_{\rm T} (\phi) \equiv \frac{1}{2}\Lambda^4{\rm tanh}^{2} \left( \frac{\phi}{M} \right) \ , \label{eq:alphaattractor}
\ee
which introduces an inflection point separating both regimes at $\phi \sim M$. In the limit $M \rightarrow \infty$ we recover the monomial function \eqref{eq:mon-pot}. Thus, we refer to Eq.~\eqref{eq:mon-pot} as the \textit{monomial limit} or \textit{approximation}. Observational constraints for the tensor-to-scalar ratio give the constraint $M \lesssim 8.5 m_{\rm pl}$, see App.~\ref{App:InflatonModel} for more details. 

In Sect.~\ref{sec:LinearAn} we will investigate the post-inflationary dynamics of our model in the monomial limit within the scope of a linearized analysis. To guarantee that the potential is bounded from below and has a single minimum at $\phi = X = 0$, we impose the condition $\lambda > h^2/(2 m_{\phi}^2)$ in our analysis \cite{Dufaux:2006ee}. Afterwards, in Sect.~\ref{Sec:Lattice} we will present a lattice-based analysis of the reheating stage in a realization of the $\alpha$-attractor T-model \eqref{eq:alphaattractor} in which the monomial approximation is valid (in the $\alpha$-attractor T-model this is guaranteed for $M \gtrsim \mpl$), which fully capture non-linear effects that become relevant at late times, and allow us to resolve the long term evolution of the energy distribution and equation of state.\footnote{Note that the stability condition $\lambda > h^2/(2 m_{\phi}^2)$ is only strictly true for the monomial potential \eqref{eq:mon-pot}, while it relaxes slightly for the realizations of the $\alpha$-attractor T-model \eqref{eq:alphaattractor} considered in Sect.~\ref{Sec:Lattice}.} Note, however, that our results from both our linearized and lattice analysis will also apply to other inflaton potentials, as long as the post-inflationary oscillations of the inflaton occur in the quadratic region of the potential. Finally, in Sect.~\ref{sec:RehCMB} we will incorporate the perturbative decay of the inflaton into our analysis by solving an appropriate set of Boltzmann equations.

Let us remark that one could naturally expand potential \eqref{eq:mon-pot} by including a `quadratic-quadratic'  (scale-free) interaction $g^2 \phi^2 X^2$ between the inflaton and the daughter field, with $g$ a dimensionless constant. This coupling does not allow for a perturbative decay of the inflaton, but it affects the details of the preheating stage: it can lead to an additional excitation of the daughter field through parametric resonance \cite{Kofman:1994rk,Shtanov:1994ce,Kaiser:1995fb,Khlebnikov:1996zt,Prokopec:1996rr,Figueroa:2017vfa}, but also generates an additional mass term that might block the preheating triggered by the trilinear interaction. In App.~\ref{App:Quadratic} we analyze the effects of such a coupling in more detail. In any case, let us here remark that the effects of a quadratic-quadratic interaction are negligible when $g^2\mpl/h \lesssim 1$, so the results presented in the bulk text (which assume $g=0$ exactly) are also valid for these cases.

\subsection{Equations of motion}

Before discussing our results, let us present the relevant equations of motion (EOM) and introduce some notation that will help us in our analysis. 

We denote $\phi_e$ as the field amplitude at the end of inflation, defined when the first slow-roll parameter obeys $\varepsilon_V (\phi_e) = 1$ (see App.~\ref{App:InflatonModel} for explicit expressions of these quantities). When inflation ends, the energy density of the universe is dominated by the homogeneous mode of the inflaton, which we denote as $\bar{\phi} = \bar{\phi} (t)$. Shortly afterwards, the inflaton enters the oscillatory regime at the amplitude $\phi_*\equiv \bar{\phi}(t_*)$. We set the scale factor to $a(t_*) = 1$ at this time from now on. In the case of the quadratic potential \eqref{eq:mon-pot}, the evolution of the homogeneous inflaton mode in this regime is approximately
\be
\bar{\phi}(t)={\mathcal A(t)}{\rm cos}(\omega_* t) \ ; \hspace{0.5cm} {\mathcal A(t)}\equiv \phi_* \left( \frac{t_*}{t} \right) , \  \label{eq:infl-hom} 
\ee
with $\omega_* \equiv m_{\phi}$ the oscillation frequency. The homogeneous inflaton oscillations give rise to a matter-dominated effective equation of state ${\bar w} = 0$ \cite{Turner:1983he}.

Given Eq.~\eqref{eq:infl-hom}, it is convenient to define a new set of `natural' dimensionless field variables $\{ \varphi,\chi \}$ and spacetime coordinates $y^{\mu} \equiv (u,\vec{y})$ as follows,
\begin{align} \varphi & \equiv \frac{1}{\phi_*} a^{3/2} \phi \ , \hspace{0.4cm} \chi \equiv \frac{1}{\phi_*} a^{3/2} X \ , \label{eq:newvars2} \\
t \rightarrow u & \equiv \int_{t_*}^t {\omega_* \,dt'} \ , \hspace{0.4cm} \vec{x} \rightarrow \vec{y} \equiv \omega_* \vec{x} \ ,\label{eq:newvars1}\end{align}
which will help us gain an analytical understanding of the post-inflationary dynamics. In these variables, the homogeneous mode (\ref{eq:infl-hom}) is just given by $\bar{\varphi} (u) = \cos (u)$, i.e.~we have reabsorbed its decaying amplitude in the new variable definitions. The EOM in terms of these new variables are 
\bea
&& \hspace{-0.4cm} \varphi'' - \frac{1}{a^2}\nabla^2_{\vec{y}}\varphi + \left( 1  + \mathcal{F}(u)\right)\varphi  + \frac{1}{2}\tilde{q}^{(h)} \chi^2 = [...] \, , \label{eq:fullEOMs1} \hspace{0.7cm}  \\
&& \hspace{-0.4cm} \chi ''- \frac{1}{a^2}\nabla^2_{\vec{y}}\chi  + \left( \tilde{q}^{(h)}\varphi + \tilde{q}^{(\lambda)} \chi^2  + \mathcal{F}(u)\right) \chi = 0 \, , \label{eq:fullEOMs2}
\eea
where $\nabla_{\vec{y}} \equiv \omega_*^{-1} \nabla$, $\mathcal{F}(u) \equiv  -3\left( a' / a \right)^2 - (3/2) a'' / a  \propto u^{-2}$ is a function that becomes subdominant very quickly, and $[...]$ denotes higher order potential terms beyond the monomial approximation, which can be neglected after the first oscillation in the T-model case if $M>\mpl $. We have also defined the \textit{effective resonance parameter} $\tilde{q}^{(h)}$ and \textit{effective self-coupling parameter} $\tilde{q}^{(\lambda)}$ as the following functions of the scale factor,
\bea
&\tilde{q}^{(h)} \equiv  q_*^{(h)} a^{-3/2} \ , \hspace{0.8cm} &q_*^{(h)} \equiv {\frac{h\phi_{*}}{ \omega_{*}^2}} \ , \label{eq:respar}\\
&\tilde{q}^{(\lambda)} \equiv  q_*^{(\lambda)} a^{-3} \ , \hspace{0.8cm} &q_*^{(\lambda)} \equiv \frac{\lambda \phi_*^2}{\omega_*^2}  \label{eq:resself}\ ,
\eea
where $q_*^{(h)}$ and $q_*^{(\lambda)}$ are dimensionless numbers proportional to the coupling strengths. At the onset of the oscillatory regime, $\{\tilde{q}^{(h)},\tilde{q}^{(\lambda)}\}$ coincide with $\{q_*^{(h)},q_*^{(\lambda)} \}$, but their values decrease over time as the universe expands. Note that we require $q_*^{(\lambda)}>(q_*^{(h)})^2/2$ in order to have a stable potential.

\begin{figure} 
    \includegraphics[width=0.48\textwidth]{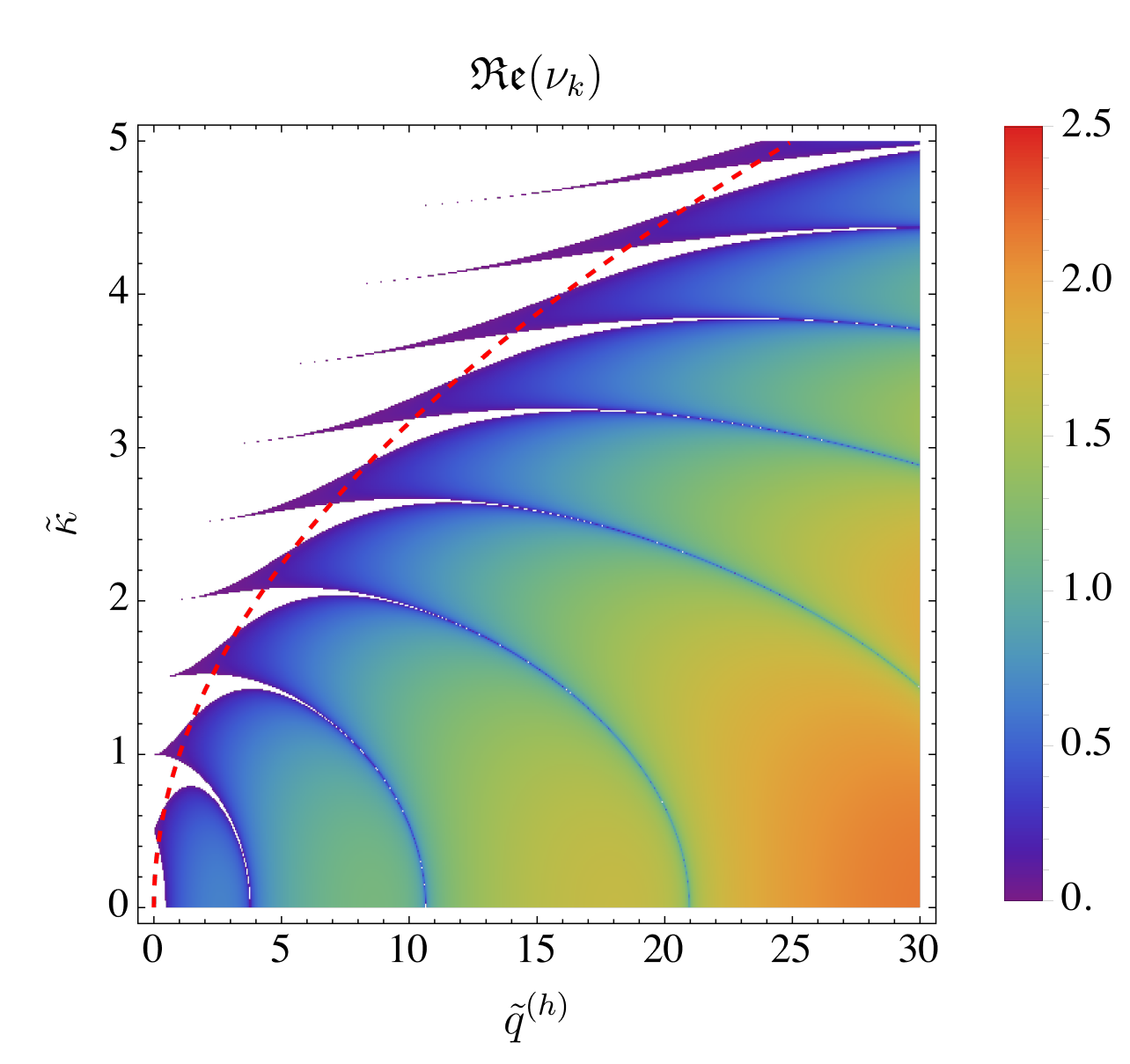}
    \caption{Stability chart for the daughter field coupled to the inflaton through a trilinear interaction. We depict the real part of the Floquet index $\nu_k$ as a function of the resonance parameter $\tilde{q}^{(h)}$ and momentum $\tilde{\kappa}$. The white area correspond to the stable region where ${\mathfrak Re}[\nu_k]=0$. The dashed red line shows the maximum momentum $\tilde{\kappa}_{\rm max} = \sqrt{\tilde{q}^{(h)}}$ experiencing tachyonic resonance for a given $\tilde{q}^{(h)}$.}
    \label{fig:FlogChart}
\end{figure}

\section{Tachyonic Resonance after Inflation: Linearised Analysis} \label{sec:LinearAn}

The early preheating stage in model (\ref{eq:model}) is dominated by a process of \textit{tachyonic resonance} for certain choices of model parameters. The trilinear interaction between the inflaton and the daughter field induces a time-dependent effective mass to the latter, which becomes periodically tachyonic due to the inflaton oscillations and leads to a strong excitation of daughter field fluctuations. This process was studied analytically in \cite{Dufaux:2006ee}. In the following we provide a semi-analytical  study of the tachyonic resonance stage and the early backreaction regime within the scope of a linearized analysis, which is valid as long as the contribution of the field fluctuations to the total energy density remains subdominant.

Let us study the process of resonant excitation in the monomial limit (\ref{eq:mon-pot}), i.e.~taking $M \rightarrow \infty$. We also neglect the $\propto \mathcal{F}(u)$ terms in equations \eqref{eq:fullEOMs1}-\eqref{eq:fullEOMs2}, since they become subdominant once the inflaton starts oscillating. In order to obtain a set of linearized evolution equations, we first expand the fields up to linear order in fluctuations, and then treat the quadratic and cubic terms by means of a Hartree-Fock approximation, i.e.~making the following substitutions in the equations of motion:~$\chi^2\rightarrow\langle\chi^2\rangle$ and $\chi^3\rightarrow3\langle\chi^2\rangle \chi$, with the variance given by $\langle \chi^2\rangle \equiv \int d {\rm log}k\, {\mathcal P}_{\chi}(k)$ (see e.g.~\cite{Antusch:2022mqv}).

The EOM of the homogeneous inflaton field $\bar{\varphi} = \bar{\varphi} (u)$ and the daughter field modes $\delta \chi_k$ are, under this approximation,\footnote{Note that in this scenario, the inflaton fluctuations are negligible except deep inside the non-linear regime, when the Hartree-Fock approximation loses its validity. We hence neglect terms $\propto \langle \varphi^2 \rangle$ in Eqs.~\eqref{eq:eomvarphi}-\eqref{eq:modechi}.}
\begin{align}
\bar{\varphi}''+ \bar{\varphi}+\frac{1}{2}\tilde{q}^{(h)}\langle \chi^2\rangle &= 0 \ , \label{eq:eomvarphi}\\
\delta \chi^{''}_{k} + (\underbrace{\tilde{\kappa}^2 + \tilde{m}_{\chi, {\rm eff}}^2}_{\equiv \, \widetilde{\omega}_{k,\chi}^2})\delta \chi_{k}&= 0 \label{eq:modechi} \ , 
\end{align}
where $\widetilde{\omega}_{k,\chi}$ is the effective frequency in natural variables, $\tilde{\kappa}\equiv\kappa/a =  k /(a \omega_*)$  with $k \equiv |\vec{k}|$ the comoving momentum, and the effective mass of the daughter field in natural variables $\tilde{m}_{\chi} \equiv m_{\chi} / \omega_*$ is,
\be
\tilde{m}_{\chi, {\rm eff}}^2 \equiv  \tilde{q}^{(h)} \bar{\varphi} + 3 \tilde{q}^{(\lambda)}\langle \chi^2\rangle \ .\label{eq:m_eff} 
\ee

By solving numerically Eqs.~\eqref{eq:eomvarphi}-\eqref{eq:m_eff} together with the Friedmann equation $H^2 = \langle \rho \rangle /(3 \mpls)$, we can study the growth of the daughter field fluctuations after inflation, incorporating also the backreaction effects from the daughter field onto the inflaton homogeneous mode in a first approximation.\\

\textbf{Resonance Process:} The amplitude of the daughter field fluctuations $\delta \chi_k$ is small at the onset of preheating, so the EOM of the homogeneous inflaton can be approximated as $\bar{\varphi}''+\bar{\varphi}\simeq0$, with solution $\bar{\varphi}\simeq{\rm cos}(u)$, and the effective mass of the daughter field reduces to $m_{\chi, {\rm eff}}^2\simeq\tilde{q}^{(h)}\bar{\varphi}$, which becomes tachyonic when $-1 \leq \bar{\varphi} < 0$, triggering a strong excitation of all modes with momenta $\tilde{\kappa}^2\lesssim \tilde{q}^{(h)}\bar{\varphi}$.

Let us ignore the quartic self-interaction of the daughter field for now, since it is small at the beginning of the resonance process. By performing a Floquet analysis of the equations of motion, we can precisely determine the range of momenta that are initially excited. In this case, Eq.~(\ref{eq:modechi}) admits solutions of the form $\delta \chi_k\sim e^{\nu_k u}$, with $\nu_k = \nu_k (\tilde{\kappa}, \tilde{q}^{(h)})$ the so-called Floquet index. Fig.~\ref{fig:FlogChart} shows the real part of $\nu_k$ as a function of $\tilde{\kappa}$ and $\tilde{q}^{(h)}$. The larger $\tilde{q}^{(h)}$ is, the larger the average Floquet exponent and the wider the range of excited momenta. The dashed line in Fig.~\ref{fig:FlogChart} indicates the maximum momentum excited by the tachyonic resonance $\tilde{\kappa}_{\rm max}= \sqrt{\tilde{q}^{(h)}}$ for a given value of $\tilde{q}^{(h)}$. Note that some modes of higher momenta are also excited, but due to parametric resonance instead, i.e.~triggered by the violation of the non-adiabaticity condition $\tilde{\omega}_{k,\chi}'/\tilde{\omega}_{k,\chi}^2\gg1$ each time the inflaton crosses the minimum of the potential (see e.g.~\cite{Kofman:1997yn}). From the Floquet chart, we can determine that the $\kappa = 0$ mode of the daughter field is only affected when $\tilde{q}^{(h)}\geq1/2$, and that there are no modes excited through tachyonic resonance for $\tilde{q}^{(h)}<1/6$.\footnote{This can be derived by using that the left border of the first instability region in Fig.~\ref{fig:FlogChart} is given by $\tilde{\kappa}^2\simeq 1/4-\tilde{q}^{(h)}/2$ \cite{MacLachlan}.}

The resonance parameter decreases with the expansion of the universe as $\tilde{q}^{(h)} = q_*^{(h)} a^{-3/2}$, so eventually $\tilde{q}^{(h)} <q_{\rm end}^{(h)} \equiv 1/6$ and the tachyonic resonance ends. We can estimate the time scale $u_{\rm end}$ when this happens as
\be 
\tilde{q}^{(h)} (u_{\rm end}) \approx q_{*}^{(h)} u_{\rm end}^{-1} = 1/6 \,\,\,\,\,\,  \longrightarrow \,\,\,\,\,\,  u_{\rm end}  \approx  6 q_{*}^{(h)} \ , 
\ee
where we have used that $a\sim u^{2/3}$. Note, however, that there is still some excitation of daughter field modes at times $u > u_{\rm end}$ around $\tilde{\kappa}\sim1/2$ due to the remnant parametric resonance.\\

\begin{figure} 
    \includegraphics[width=0.47\textwidth]{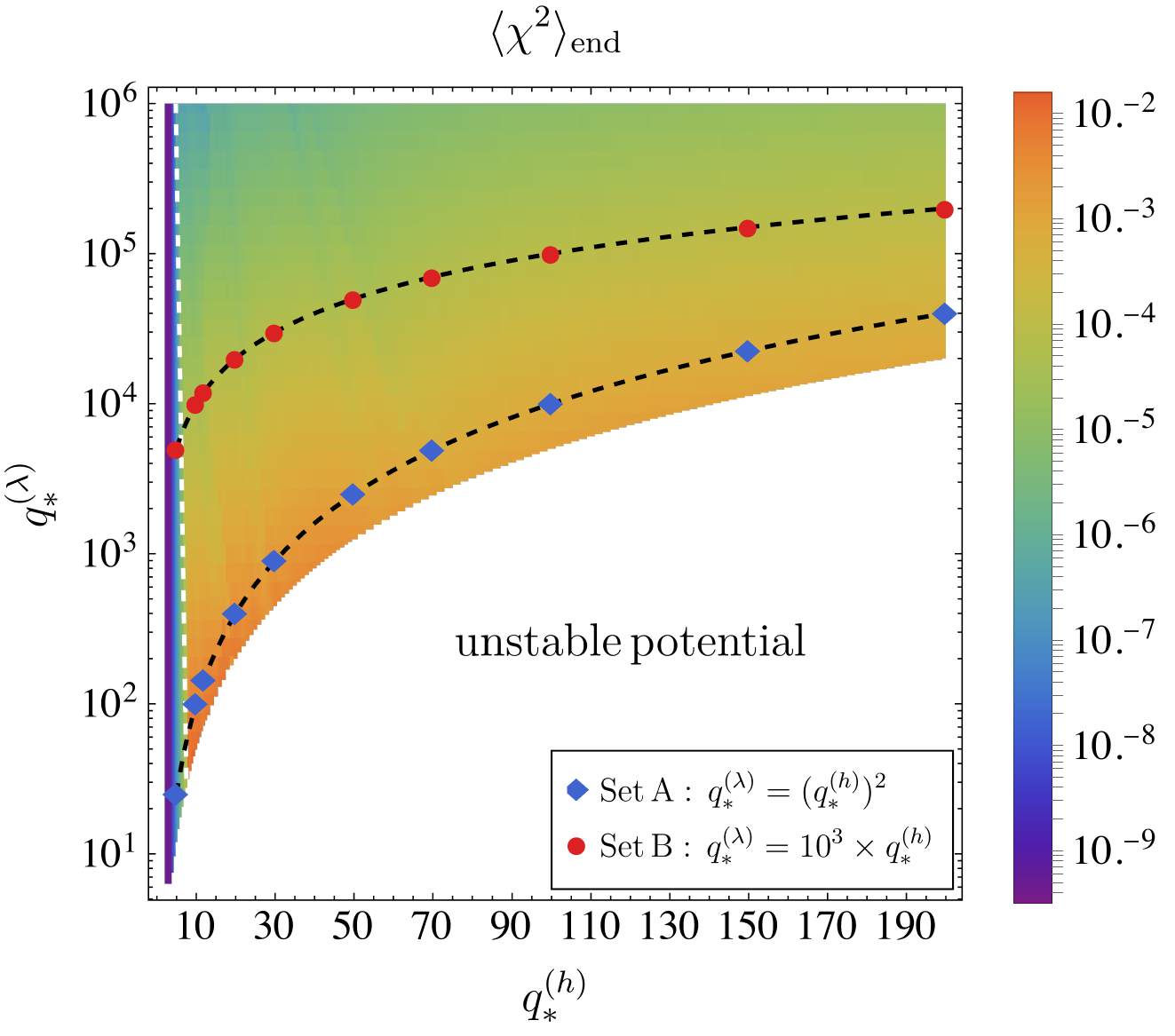}
    \caption{Value of the daughter field's variance $\langle \chi^2 \rangle_{\rm end}$ when the tachyonic resonance ends, obtained by solving Eqs.~\eqref{eq:eomvarphi} and \eqref{eq:modechi} for different choices of
    $q_*^{(h)}$ and $q_*^{(\lambda)}$. The white dashed line indicates the value of $q_{\rm min}^{(h)}$ separating the two regimes described in the main text. The white area depicts the model parameters for which the potential is unstable. The blue diamonds and red circles indicate the two sets of cases simulated in the lattice, see Sect.~\ref{sec:RehCMB} for more details.}
    \label{fig:HartreeVarMax1}
\end{figure}

\textbf{End of resonance:} In the above analysis we have ignored the quartic self-interaction of the daughter field. To analyze its impact in the tachyonic resonance process, we have solved numerically Eqs.~\eqref{eq:eomvarphi}-\eqref{eq:m_eff} together with the Friedmann equations. In the Hartree-Fock approximation, any backreaction effects from the daughter field onto the inflaton are encoded in its variance, and transmitted through the second term of Eq.~\eqref{eq:m_eff}.

\begin{figure*} 
    \includegraphics[width=0.48\textwidth]{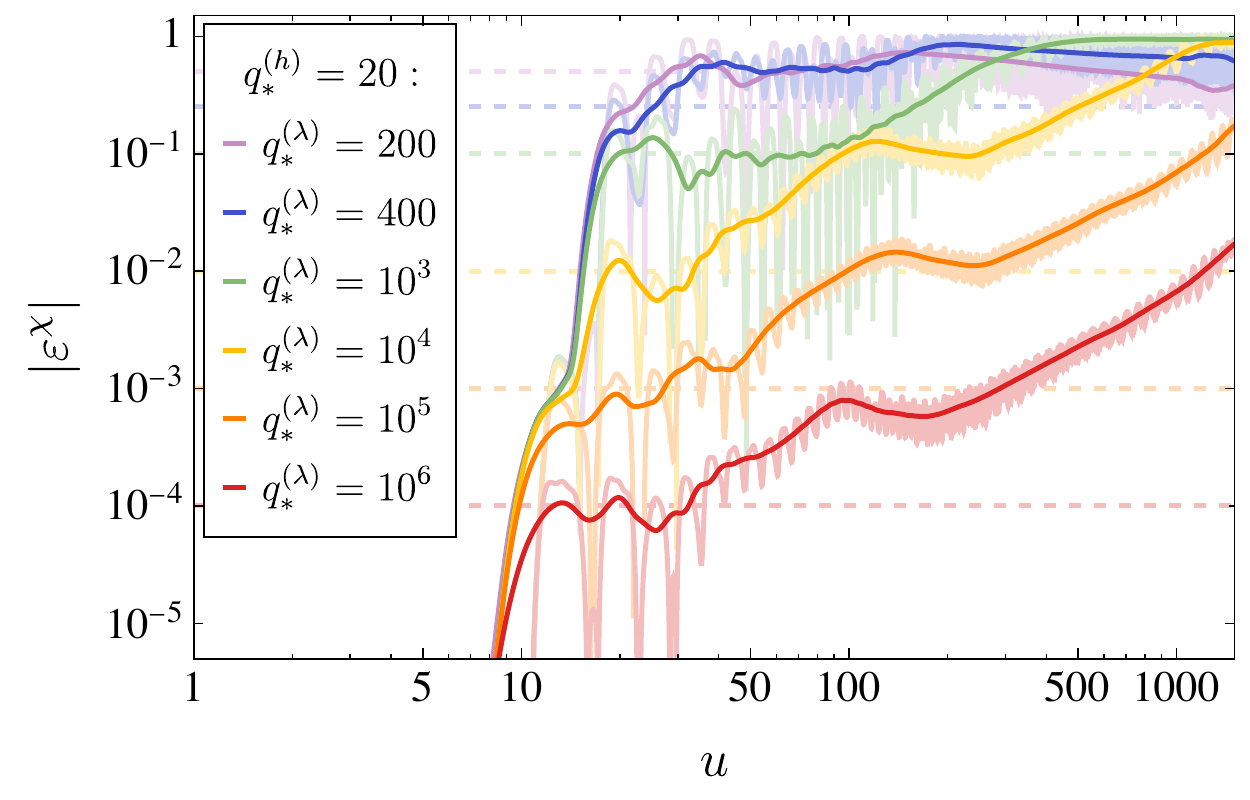}
    \includegraphics[width=0.48\textwidth]{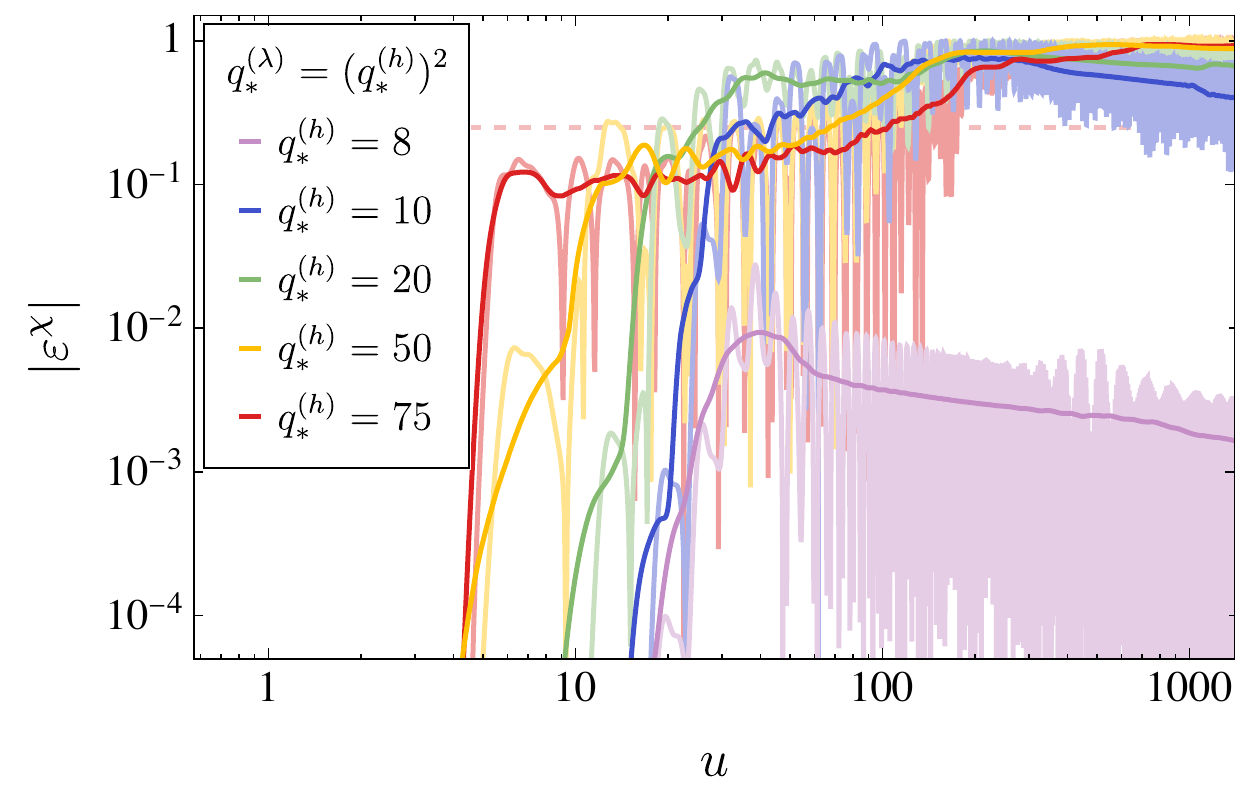}
    \caption{Evolution of the fraction of energy density stored in the daughter field $|\varepsilon^\chi|$ (light colors) and its oscillation average $\overline{|\varepsilon^\chi|}$ (dark colors). We depict the absolute value of each quantity. The left panel depicts cases for fixed $q_*^{(h)}=20$ and different values of $q_*^{(\lambda)}$, while the right panel shows cases for different $q_*^{(h)}$ and the condition $q_*^{(\lambda)}=(q_*^{(h)})$. The dashed horizontal lines indicate the estimate of $|\varepsilon^\chi|$ at the end of the tachyonic resonance, obtained with Eq.~\eqref{eq:varend} (note that in the right panel all dashed horizontal lines overlap).}
    \label{fig:lin_en}
\end{figure*} 

We have solved the equations for different choices of $q_*^{(h)}$ and $q_*^{(\lambda)}$, always obeying the stability condition $q_*^{(\lambda)}\geq (q_*^{(h)})^2/2$. Depending on the value of $q_*^{(h)}$, we have found two characteristic regimes, in which the tachyonic resonance terminates due to different reasons, and hence leads to different post-inflationary evolutions. Both regimes are separated by the critical value $q_{\rm min}^{(h)} \approx 6 - 9$, indicated by the white dashed line in Fig.~\ref{fig:HartreeVarMax1}:\footnote{In our analysis we have observed that the exact value of $q_{\rm min}^{(h)}  \approx 6 - 9$ depends slightly on the choice of $q_*^{(\lambda)}$.}

\begin{itemize}[left=0pt, labelindent=3pt]
\item[{\bf a)}] $\pmb{ q_*^{(h)} < q_{\rm min}^{(h)}}${\bf :} In these cases, the tachyonic resonance ends before the amplitude of the daughter field modes becomes sizable, i.e.~the oscillating homogeneous mode of the inflaton remains the dominant energy contribution at all times, and the deviation of the equation of state from $\bar \omega = 0$ is minimal.

\item[{\bf b)}] $\pmb{q_*^{(h)}>q_{\rm min}^{(h)} }${\bf :} In these cases, the daughter field variance grows exponentially until the term $\propto \tilde{q}^{(\lambda)}\langle \chi^2\rangle$ in (\ref{eq:modechi}) becomes large enough for the daughter field's effective mass to stop being tachyonic, shutting off the resonance. The daughter field variance when the tachyonic resonance ends $\langle \chi^2\rangle_{\rm end}$, obtained numerically, is depicted in Fig.~\ref{fig:HartreeVarMax1}. The value of $\langle \chi^2\rangle_{\rm end}$ can also be roughly estimated as follows, 
\be \langle \chi^2\rangle_{\rm end} \approx \frac{\tilde{q}^{(h)}\bar{\varphi}}{3\tilde{q}^{(\lambda)}} \sim \frac{q_*^{(h)}}{3 q_*^{(\lambda)}} a_{\rm end}^{3/2} \ \label{eq:varend} ,\ee
where $a_{\rm end}$ denotes the scale factor at that time. In our numerical computation we observe that the tachyonic resonance stage may terminate very fast: for $q_*^{(h)}\gtrsim75$, it terminates after less than one inflaton oscillation, while for $30 \lesssim q_*^{(h)} \lesssim 75 $ it takes less than two oscillations. Remarkably, this happens when the field fluctuations are still energetically subdominant, i.e.~the energy density remains dominated by the oscillating inflaton homogeneous mode. However, $\tilde{q}^{(\lambda)}$ continues decreasing due to the expansion of the universe, which allows the daughter field variance to continue growing, albeit at a much slower rate than during the previous stage of tachyonic resonance. This can be clearly seen in Fig.~\ref{fig:lin_en}, which depicts the evolution of the fraction of energy density stored in the daughter field $\varepsilon^\chi$ (defined explicitly in Eq.~\eqref{eq:linEchi} of App.~\ref{App:EnergyEos}) 
for different choices of $q_*^{(h)}$ and $q_*^{(\lambda)}$. We can see that it eventually grows up to $\bar{\varepsilon}^\chi \sim 10^{-1} - 1$ for all resonance parameters $q_*^{(h)}>q_{\rm min}^{(h)}$. Note also that, due to the forcing term in (\ref{eq:eomvarphi}), the inflaton develops a vacuum expectation value $v \simeq -\tilde{q}^{(h)}\langle \chi^2\rangle/2$ at late times.

\end{itemize}

\section{Tachyonic resonance after inflation: Lattice Analysis}\label{Sec:Lattice}

The linearized analysis presented in the previous section is well suited to study the early post-inflationary dynamics, but does not suffice once non-linear effects become relevant. In this section we present results from lattice simulations carried out with the code \CL \cite{Figueroa:2021yhd}, which fully capture these non-linear effects. As the system usually takes very long to equilibrate after preheating, we have decided to run simulations in 2+1 dimensions with an appropriately modified version of the public code. In App.~\ref{App:3D} we compare results of simulations performed in 2+1 and 3+1 dimensions, which supports the suitability of this technique.

More specifically, we simulate the post-inflationary dynamics of model (\ref{eq:model}) with the inflaton potential given by the $\alpha$-attractor T-model (\ref{eq:alphaattractor}). We have chosen $M=5\mpl$, for which the monomial approximation applies. We set the scale of inflation to $\Lambda\simeq0.0056\mpl$, which corresponds to $N_k = 55$, see App.~\ref{App:InflatonModel} for more details. The fields are initialized at the end of inflation, with their homogeneous modes given by $\bar{\phi}=\phi_{\rm e}\simeq1.348 \mpl$ and $\bar{\chi}=0$, over which we impose a distribution of vacuum fluctuations. We have simulated the system in regular lattices of $256$ and $512$ points per dimension, and with \textit{velocity-verlet} integrators of up to fourth order of accuracy, which guarantee sufficient energy conservation at late times.

In our analysis we study the evolution of the energy ratios $\varepsilon_{\alpha}$, defined as the relative contribution of each energy density component to the total. These include the kinetic and gradient contributions of each field ($\varepsilon_{\rm k}^{\varphi}$, $\varepsilon_{\rm g}^{\varphi}$, $\varepsilon_{\rm k}^{\chi}$ and $\varepsilon_{\rm g}^{\chi}$), and the three potential contributions ($\varepsilon_{\rm p}^{\varphi}$, $\varepsilon_{\rm i}$ and $\varepsilon_{\rm p}^{\chi}$), corresponding to three terms of potential \eqref{eq:model}. Explicit definitions for these quantities are provided in App.~\ref{App:EnergyEos}. We also study the evolution of the equation of state $w$ and its \textit{effective} (i.e.~oscillation-averaged) evolution $\bar{w}$. The equation of state can be written as 
\be w \equiv \frac{p}{\rho} = \varepsilon_{\rm k}  -\frac{1}{3} \varepsilon_{\rm g} - \varepsilon_{\rm p} \label{eq:eosEnRat}  \ , \hspace{0.5cm} \left[ \begin{array}{l}
    \varepsilon_{\rm k} \equiv \varepsilon_{\rm k}^\varphi+\varepsilon_{\rm k}^\chi \\
    \varepsilon_{\rm g} \equiv \varepsilon_{\rm g}^\varphi+\varepsilon_{\rm g}^\chi \\
    \varepsilon_{\rm p} \equiv \varepsilon_{\rm p}^\varphi+\varepsilon_{\rm p}^\chi + \varepsilon_{\rm i}
\end{array} \right] \ , \ee
where the RHS must be interpreted as a volume average over the lattice. The effective equation of state $\bar{w}$ is sourced instead by the oscillation-average ratios $\bar{\varepsilon}_{\alpha}$.

\subsection{Evolution of the energy distribution and the equation of state}

\begin{figure*}
    \includegraphics[width=0.46\textwidth]{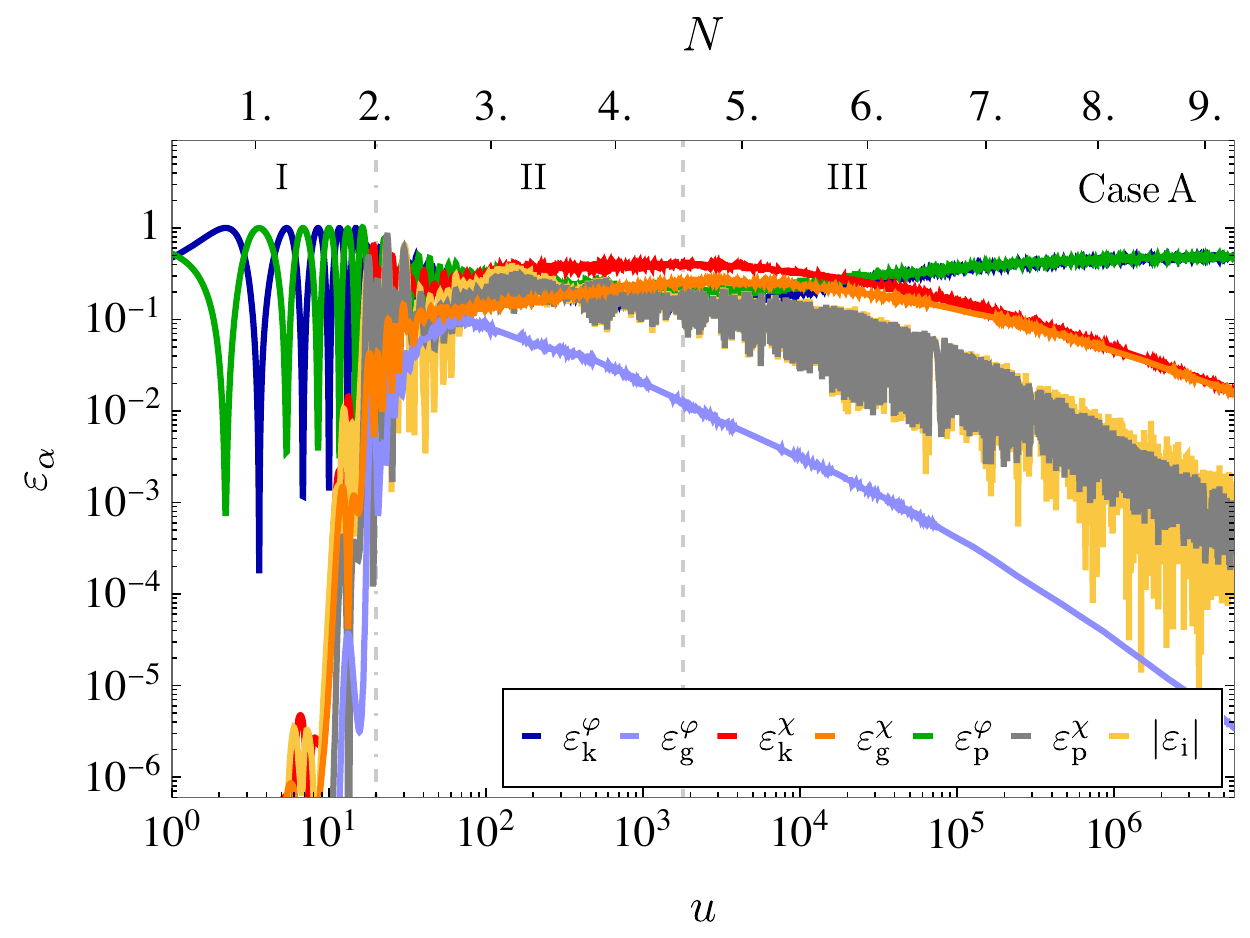} \hspace{0.3cm} 
    \includegraphics[width=0.46\textwidth]{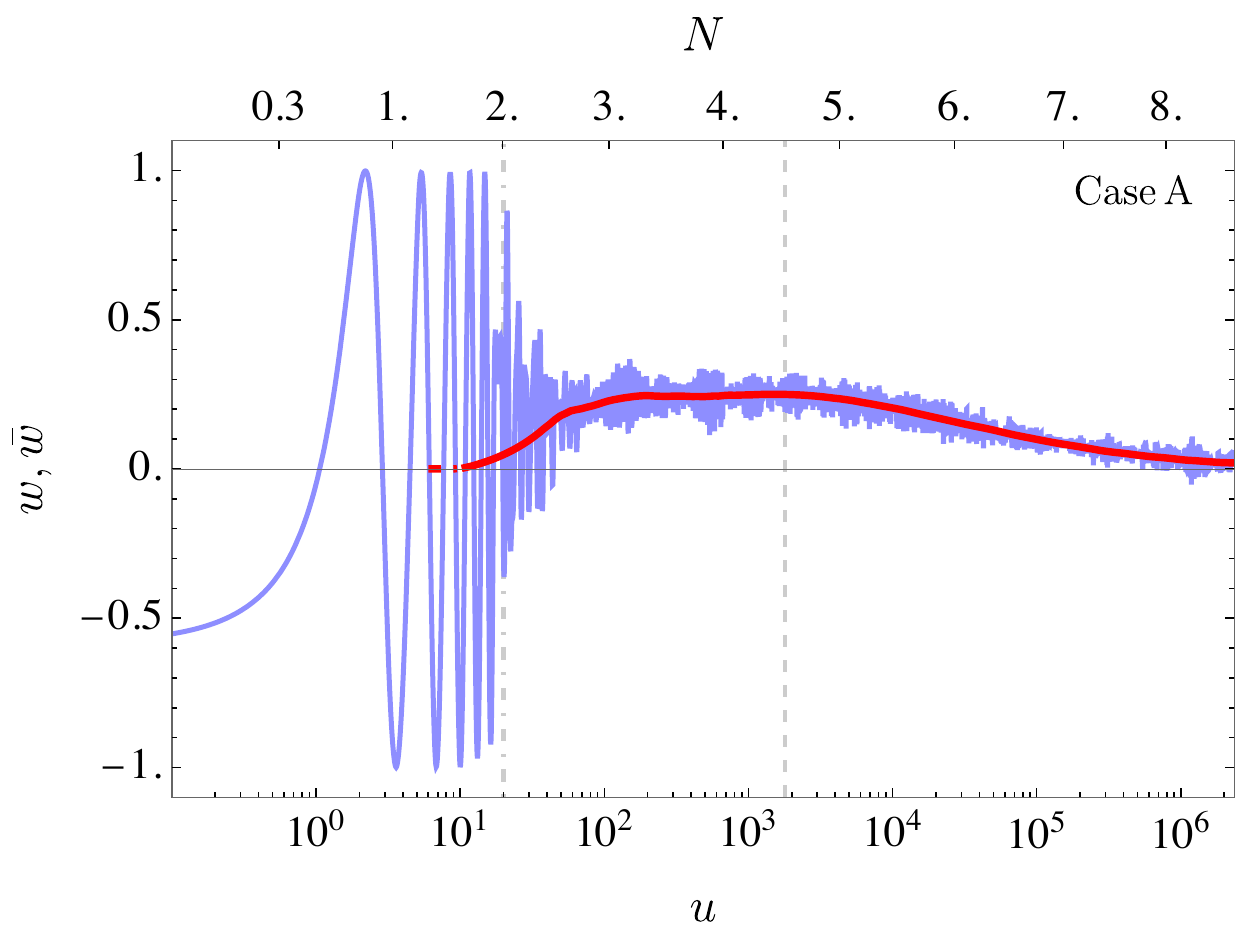}\\ \vspace{0.5cm}
    \includegraphics[width=0.46\textwidth]{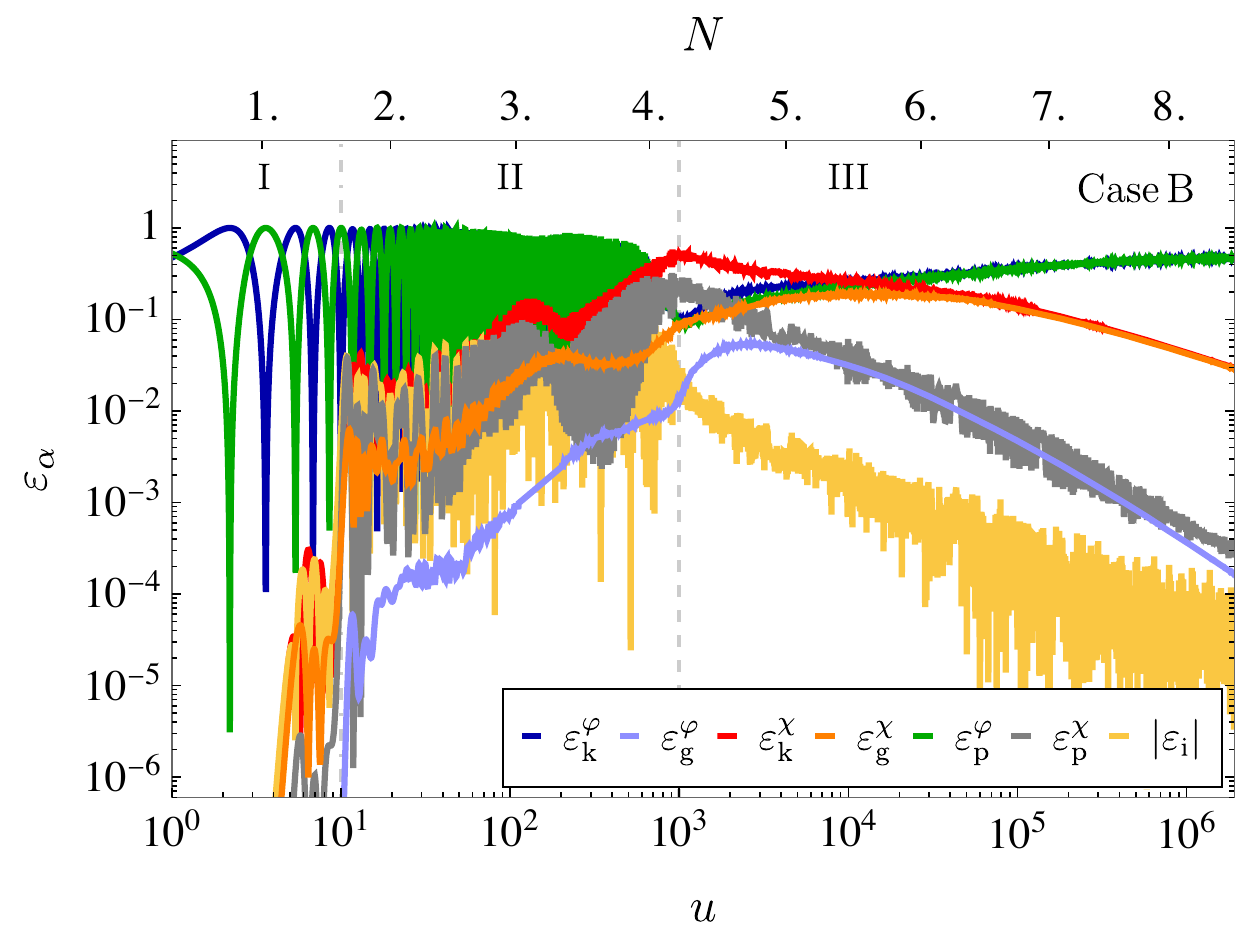} \hspace{0.3cm} 
    \includegraphics[width=0.46\textwidth]{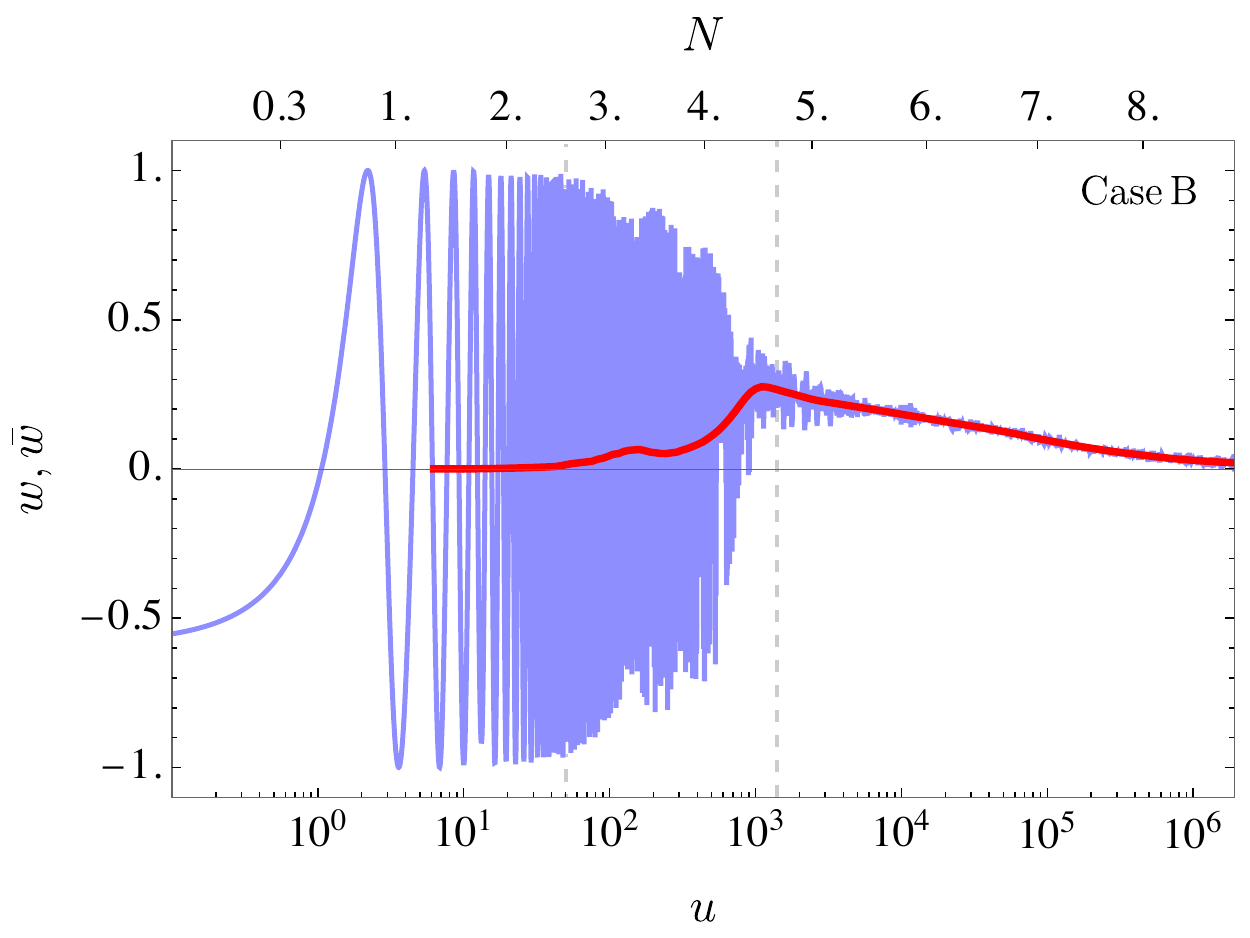}
    \caption{Left panels: Evolution of the energy ratios $\varepsilon_{\alpha}$ for Case A (top) and Case B (bottom) as a function of time $u$ and number of e-folds, obtained from 2+1-dimensional lattice simulations. Note that for the interaction energy ratio we plot the absolute value $|\varepsilon_i|$, as it can become negative. The vertical lines in each panel delimit Stages I, II and III of the post-inflationary evolution, described in the bulk text. Right panels: Evolution of the instantaneous equation of state $w$ (blue) and its oscillation-average $\bar{w}$ (red) for the same cases.}
    \label{fig:EnDenTri}
\end{figure*}

We consider the following two choices of coupling parameters, which we refer to as as Case A and Case B:
\begin{itemize}
    \item {\bf Case A}: \hspace{0.4cm} $q_*^{(h)}=50$, \, $q_*^{(\lambda)}= 2500$\ ,
    \item {\bf Case B}: \hspace{0.36cm} $q_*^{(h)}=50$, \, $q_*^{(\lambda)} = 50000$\ .
\end{itemize}
These correspond to fixing $h=1.5 \cdot 10^{-9} \mpl$ in both cases, as well as $\lambda=5.5 \,\cdot\, 10^{-8}$ and $1.1\cdot 10^{-6}$ in Case A and B respectively, see App.~\ref{sec:CoupConsts}.

In the left panels of Fig.~\ref{fig:EnDenTri} we show the evolution of the energy ratios in both cases, while the right panels show the evolution of both the instantaneous and effective equation of state. We can divide the preheating phase into three stages:\\

{\bf Stage I ($\pmb{\bar{w}\approx0}$):} Immediately after inflation, the energy density is dominated by the homogeneous mode of the inflaton, with oscillation-averaged energy ratios ${\bar{\varepsilon}_{\rm k}^{\varphi} \simeq \bar{\varepsilon}_{\rm p}^{\varphi} \simeq 1/2}$. The instantaneous equation of state oscillates in the interval $-1 \leq w \leq +1$, with the oscillation average yielding $\bar{w}\simeq 0$. Although the daughter field fluctuations are energetically subdominant during this stage, their amplitude grows exponentially due to their tachyonic excitation in both cases, in agreement with the linearized analysis of Sect.~\ref{sec:LinearAn}. The inflaton fluctuations are also amplified during this stage due to non-linear backreaction effects from the daughter field. \\

{\bf Stage II ($\pmb{\bar{w}\rightarrow \bar{w}_{\rm max}}$):} After few oscillations, the amplitude of the field fluctuations has grown so much that the inflaton homogeneous mode decays. Correspondingly, the gradient energy ratios become sizable, and $\bar{w}$ begins to significantly deviate from $\bar{w} = 0$, eventually reaching a local maximum $\bar{w} = \bar{w}_{\rm max} < 1/3$. However, the detailed evolution of the energy ratios and equation of state during this stage proceeds very differently in the two cases, due to the different strengths of the daughter field self-interaction:

\begin{itemize}
\item In Case A, the self-interaction is weaker and does not significantly dampen the tachyonic resonance, so inflaton fragmentation occurs very quickly, and the equation of state attains its maximum value $\sim 3$ e-folds after inflation ends. Approximately at this time, backreaction effects freeze out and $\varepsilon_{\rm g}^{\varphi}$ starts decreasing. However, due to non-linear effect emerging from the quartic self-interaction of the daughter field, energy is efficiently redistributed to the daughter field, which in turn causes an increase of $\varepsilon_{\rm g}^{\chi}$ (this phenomenon is similar to the one found in \cite{Antusch:2022mqv} for $\phi^2 X^2$ interactions). Both effects compensate in such a way that the sum $\varepsilon_{\rm g} = \varepsilon_{\rm g}^{\varphi} + \varepsilon_{\rm g}^{\chi}$ remains constant, and hence the equation of state remains at $\bar{w} \approx \bar{w}_{\rm max}$ for several e-folds, developing a plateau.

\item In Case B, the quartic self-interaction is much stronger, which leads to an early termination of the exponential growth of field fluctuations triggered by the tachyonic resonance. However, in agreement with the linearized analysis of Sect.~\ref{sec:LinearAn}, the daughter field energy ratio continues growing, albeit at a much slower rate than before. Thus, the fragmentation of the homogeneous inflaton mode happens more gradually than in Case A, and the equation of state attains its maximum value $\bar{w}_{\rm max}\approx0.275$ at later times relative to Case A.
\end{itemize}

Note also that in both cases, due to the forcing term $\propto \tilde{q}^{(h)}\chi^2$ in Eq.~(\ref{eq:fullEOMs1}), the inflaton daughter field variance roughly follows the relation $\langle \chi^2 \rangle\approx2|\bar{\varphi}|/\tilde{q}^{(h)}$ after the equation of state reaches its maximum, in agreement with our linearized analysis and with \cite{Dufaux:2006ee}.\\

{\bf Stage III ($\pmb{\bar{w}\rightarrow 0}$):} At very late times, $\tilde{q}^{(h)},\,\tilde{q}^{(\lambda)} \ll 1$, and fluctuations are no longer efficiently excited. From that moment on, the inflaton and daughter field fluctuations dilute as radiation, $\rho_\chi,\,\rho_{\delta \phi} \sim a^{-4}$, while the inflaton homogeneous mode behaves as matter, $\rho_{\bar{\phi}} \sim a^{-3}$. Hence, the inflaton gradually recovers a larger fraction of the total energy density as the system continues evolving.
In turn, the effective equation of state decreases from its maximum value $\bar{w}_{\rm max}$ towards $\bar{w} \rightarrow 0$. Eventually, the energy density is dominated again by the inflaton's oscillating homogeneous mode.\newline

Note that in both cases we only observe a partial fragmentation of the inflaton. Full fragmentation would occur if the energy ratios were $\bar{\varepsilon}_{\rm k} = \bar{\varepsilon}_{\rm g} \simeq 1/2$ and $\bar{\varepsilon}_{\rm p}\simeq 0$, yielding $\bar{w} = 1/3$ through Eq.~\eqref{eq:eosEnRat}. In summary, preheating on its own neither successfully depletes the inflaton energy nor produces a radiation-dominated universe. Reheating is instead achieved perturbatively at a later stage, as discussed in the following section.

\section{Perturbative reheating} \label{sec:RehCMB}

\begin{figure*}
    \includegraphics[width=0.48\textwidth]{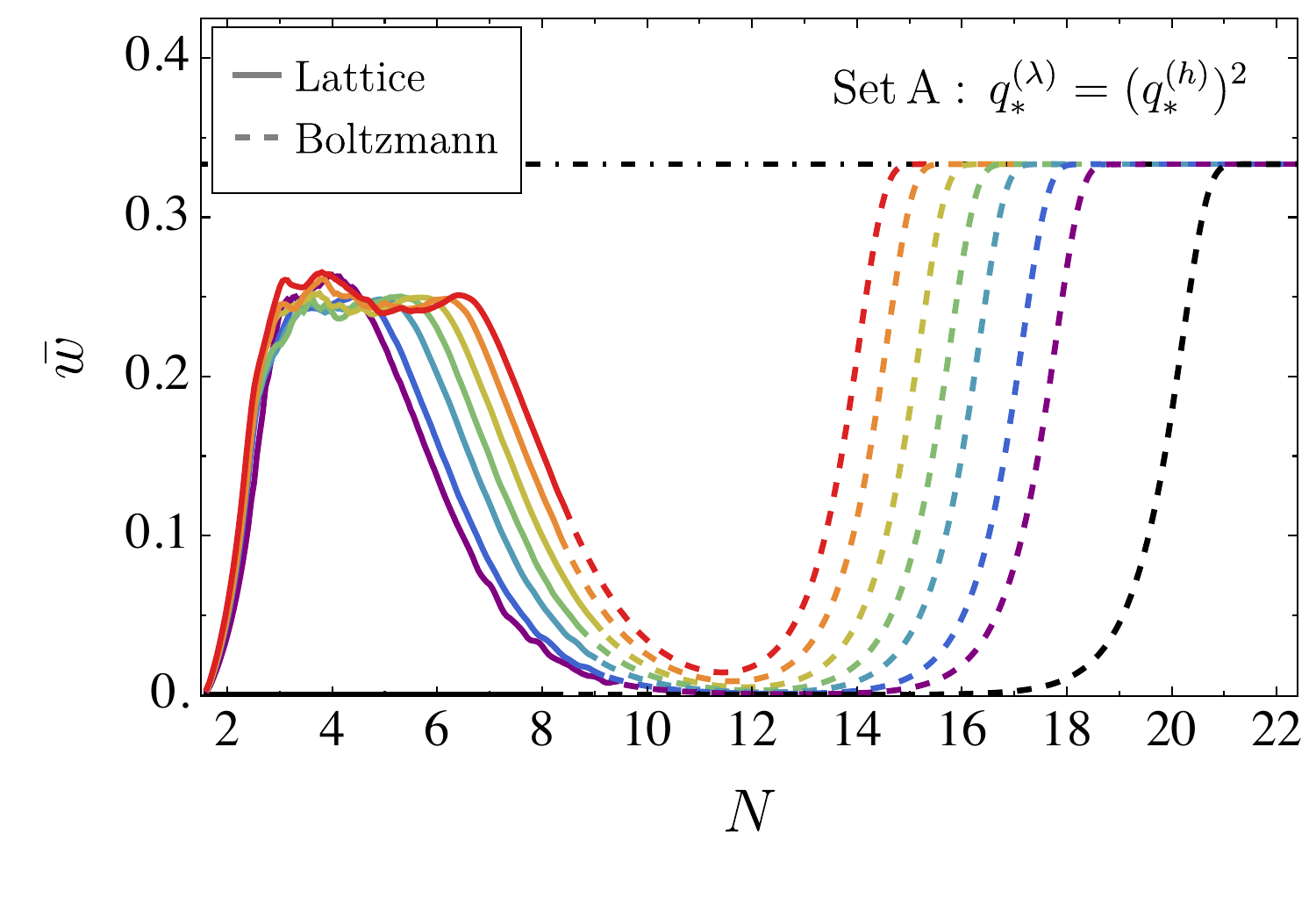}
     \includegraphics[width=0.48\textwidth]{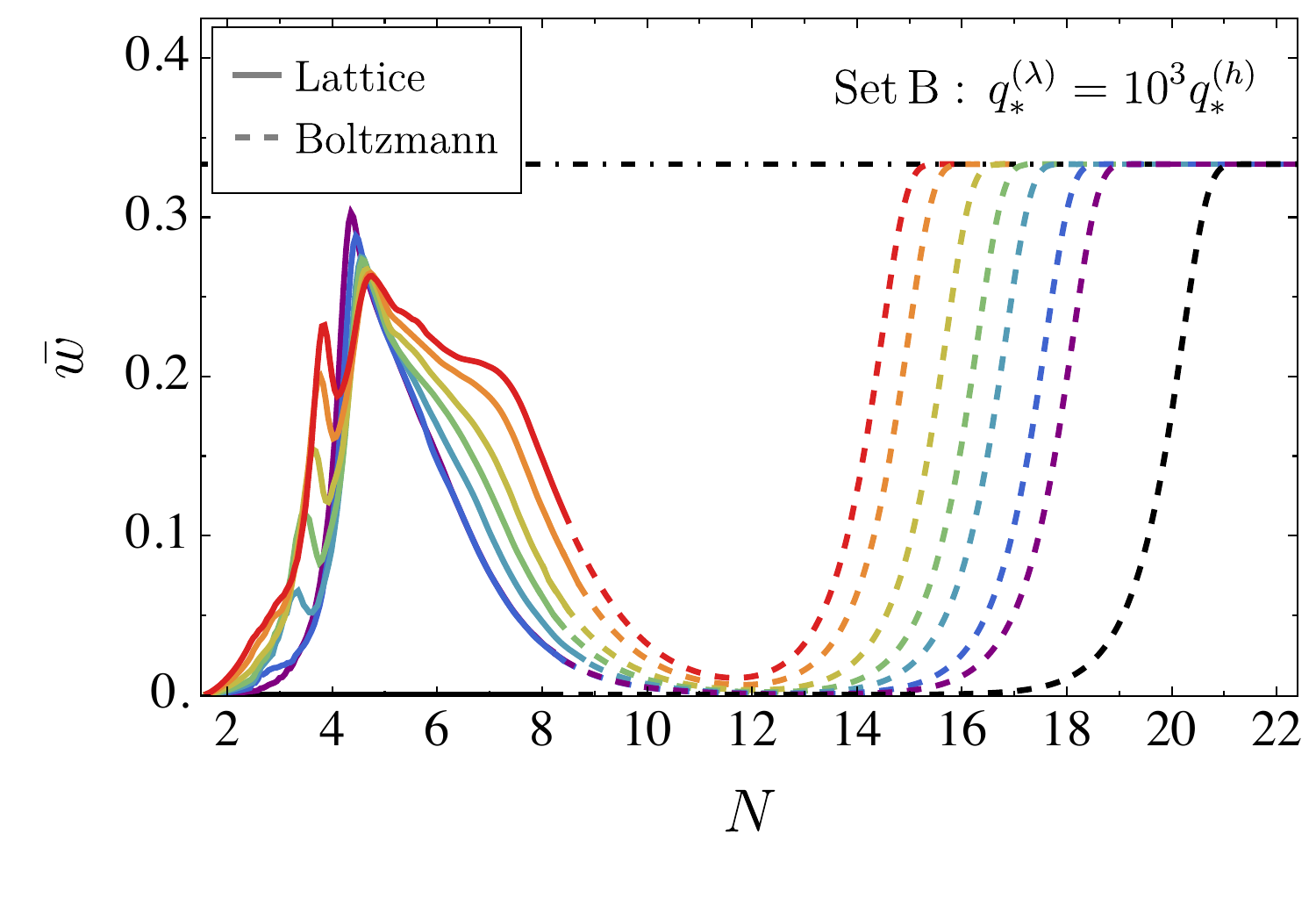}
     \includegraphics[width=0.75\textwidth]{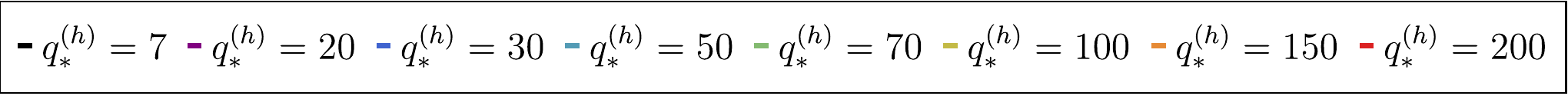}
    \caption{Evolution of the effective equation of state as a function of number of e-folds $N$ from the end of inflation until the system reaches a radiation dominated state, for different parameter choices in Set A (left) and Set B (right). The solid lines show the initial evolution obtained from the 2+1-dimensional lattice simulations, while the dashed lines show the later evolution obtained by solving the Boltzmann equations \eqref{eq:BE1}-\eqref{eq:BE3}. The horizontal dotted-dashed line indicates $\bar{w}_{\rm rd}=1/3$.}\label{fig:eos_full}
\end{figure*} 

In the previous sections we have discussed the details of the preheating stage. We now focus on the later evolution until the inflaton decays perturbatively and the system becomes fully radiation dominated. The trilinear interaction makes the inflaton decay into two $\chi$-field quanta with the following decay rate, 
\be
\Gamma_\phi=\frac{h^2}{32\pi m_\phi}  \ ,\label{eq:decayRate}
\ee
which becomes effective when $H\approx \Gamma_\phi$. This condition occurs much later than the final time captured by our lattice simulations, so in order to characterize the remaining reheating stage, we switch to an effective picture, in which we solve the Boltzmann equations for the inflaton and daughter field energy densities, $\rho_\varphi$ and $\rho_\chi$, together with the first Friedmann equation,\footnote{Note that the interaction energy is negligible at the end of our lattice simulations, which allows us to treat $\rho_\varphi$ and $\rho_\chi$ as two independent variables in our Boltzmann approach.}
\begin{align}
\dot{\rho}_\varphi&+3H\rho_\varphi+\Gamma_\phi \rho_\varphi=0 \ , \label{eq:BE1}  \\ 
\dot{\rho}_\chi&+4H\rho_\chi-\Gamma_\phi \rho_\varphi=0 \ ,\label{eq:BE2}  \\
H^2&=\frac{1}{3\mpl^2}(\rho_\varphi + \rho_\chi) \ . \label{eq:BE3} 
\end{align}
Here we treat the inflaton and daughter field as matter and radiation fluids respectively, since their individual equations of state are $\bar{w}_\varphi\approx0$ and $\bar{w}_\chi\approx1/3$  at the end of the simulations. As initial conditions for $\rho_\varphi$ and $\rho_\chi$, we extract their oscillation-averaged values from the lattice at the last simulated time, and then solve Eqs.~\eqref{eq:BE1}-\eqref{eq:BE3} numerically until the system becomes fully radiation dominated. \newline

{\bf Equation of state of reheating:} In the following we consider two sets of coupling parameters, which allow us to explore different regimes of parameter space and learn some generic features of how the equation of state evolves:
\begin{itemize}
    \item {\bf Set A:} \hspace{0.36cm} $q_*^{(\lambda)} = (q_*^{(h)})^2$ \hspace{0.02cm}, \hspace{0.1cm} $q_*^{(h)} \in [ 5, 200]$ \ ,
    \item {\bf Set B:} \hspace{0.33cm} $q_*^{(\lambda)} = 10^3 q_*^{(h)}$ \hspace{0.02cm}, \hspace{0.1cm} $q_*^{(h)} \in[5,200]$ \ .
\end{itemize}
The specific coupling parameters simulated are indicated in Fig.~\ref{fig:HartreeVarMax1}. Note that Cases A and B studied in the previous section belong to Set A and B respectively. 

Fig.~\ref{fig:eos_full} depicts the full evolution of the effective equation of state for several cases in Set A (left panel) and Set B (right panel), combining both the lattice output and the solution to the Boltzmann equations. Remarkably, the evolution of $\bar{w}$ follows a similar pattern in all cases of Set A: after inflaton fragmentation 2-3 e-folds after the end of inflation, the equation of state stabilizes during several e-folds around the same value $\bar{w}\simeq 0.25$, with the duration of such a plateau being longer for larger values of $q_*^{(h)}$ and $q_*^{(\lambda)}$. As explained in the previous section, the interactions eventually become inefficient due to the expansion of the universe, and the effective equation of state returns to $\bar{w} \rightarrow 0$. On the other hand, in all cases of Set B, the self-interaction of the daughter field is stronger, which terminates the initial stage of tachyonic resonance early. This is followed by a more gradual increase of $\bar{w}$, which reaches its local maximum at later times relative to Set A. Note that for the larger values of $q_*^{(h)}$ in Set B, $\bar{w}$ starts developing a plateau similar to the one in Set A.

Once the interactions have become negligible, $\bar{w}$ decreases in all cases of both sets until reaching the local minimum $\bar{w}_{\rm min}\simeq \Gamma_\phi/(3H_{\rm min})$ at $N\approx 15 (q_*^{(h)})^{-0.05}$ e-folds after the end of inflation.\footnote{At the minimum $\bar{w} = \bar{w}_{\rm min}$, we have found that the ratio of the decay rate to the Hubble parameter for all $q_*^{(h)}>q_{\rm min}^{(h)}$ is well approximated by the following expressions,
\begin{align}
\text{Set A: } \,\,\, \Gamma_\phi/H_{\rm min} & \approx 4\cdot10^{-4}(q_*^{(h)} /10)^{8/5} \ , \\ 
\text{Set B: } \,\, \Gamma_\phi/H_{\rm min} & \approx 7\cdot10^{-4}+10^{-4} ( q_*^{(h)} /10 )^{19/2} \ .
\end{align}
} At this time, the perturbative decay  starts becoming effective in comparison to the expansion rate, leading to a renewed transfer of energy from the inflaton condensate into radiation. As a result, the averaged equation of state grows again and eventually reaches $\bar{w} \rightarrow 1/3$.

\begin{figure}
    \includegraphics[width=0.47\textwidth]{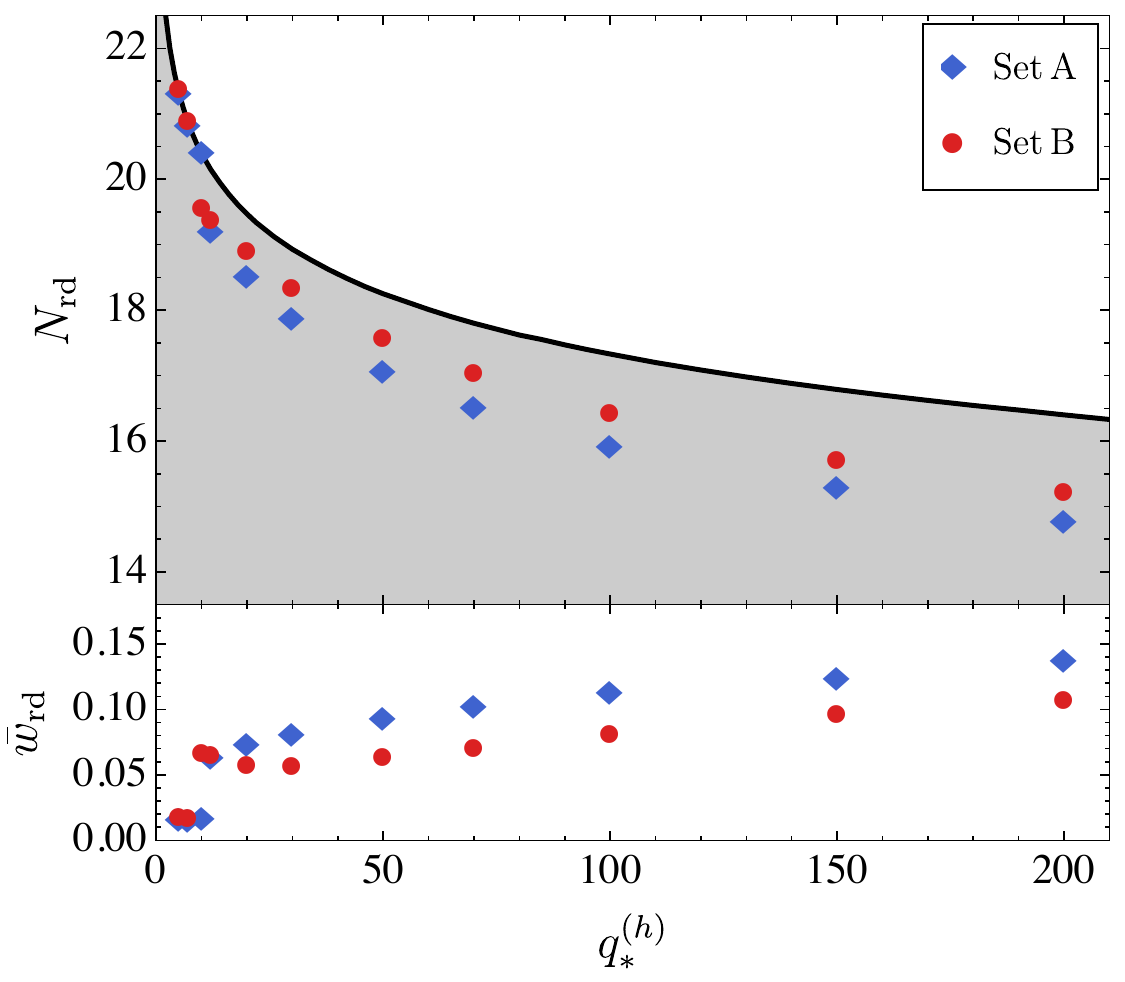}
    \caption{Number of e-folds from the end of inflation until radiation domination $N_{\rm rd}$ (top panel) and average equation of state during that period $\bar{w}_{\rm rd}$ (bottom panel) for the considered coupling parameters. The gray area depicts the theoretical uncertainty when minimal assumptions are made: the solid black line assumes a reheating phase that ignores any effects from preheating, while the lower limit corresponds to instantaneous reheating $N_{\rm rd}=0$.}
    \label{fig:Nrh}
\end{figure}

In Fig.~\ref{fig:Nrh} we show the number of e-folds from the end of inflation until the achievement of radiation domination $N_{\rm rd}$ for the considered cases, as well as the averaged equation of state during that period $\bar{w}_{\rm rd}$, defined as,\footnote{In practice, we compute $\bar{w}_{\rm rd}$ by inverting the relation $N_{\rm rd}=-(3\bar{w}_{\rm rd}+3)^{-1}{\rm ln}(\rho_{\rm rd}/\rho_{\rm end})$.}
\be \bar{w}_{\rm rd} = \frac{1}{N_{\rm rd}} \int_0^{N_{\rm rd}} d N' w (N')\ .
\ee
In this computation, radiation domination is defined as the time when radiation represents 99\% of the total energy density. The gray area in Fig.~\ref{fig:Nrh} represents the region of theoretical uncertainty of $N_{\rm rd}$ when only minimal assumptions are made about the reheating phase: the upper bound corresponds to a matter-dominated reheating phase that ends with the perturbative decay of the inflaton and ignores any relevant effects from the preheating stage, while the lower limit corresponds to an instantaneous reheating scenario $N_{\rm rd}=0$. For cases $q_*^{(h)}<q_{\rm min}^{(h)}$, the values of $N_{\rm rd}$ match the upper bound because, as explained in Sect.~\ref{sec:LinearAn}, the stage of tachyonic resonance is too short to produce relevant amounts of fluctuations, and the homogeneous inflaton mode dominates the entire reheating phase.\footnote{Note that in these cases, $\bar{w}_{\rm rd}$ is not exactly zero, because the transitions of the equation of state at the end of inflation and the onset of perturbative reheating are not instantaneous.} On the other hand, for parameters $q_*^{(h)}\gtrsim q^{(h)}_{\rm min}$ we find that, due to the preheating stage, the value of $N_{\rm rd}$ is lower by $\sim 1 - 2$ e-folds with respect to the upper bound, while $\bar{w}_{\rm rd} \sim 0.05 - 0.15$. As expected, the larger the value of $q_*^{(h)}$, the stronger the tachyonic resonance, and hence the larger the deviation. However, a simultaneous increase in $q_*^{(\lambda)}$ does not necessarily lead to a larger deviation in both quantities, as the depicted results from Set A and B show.

\section{CMB observables} \label{sec:CMBconst}

Our results on the equation of state allow us to obtain accurate predictions on the inflationary CMB observables, i.e.~the spectral index $n_s$ and the tensor-to-scalar ratio $r$, and assess how these are impacted by the preheating stage. These observables depend on the number of e-folds of expansion from the time the pivot scale $k_{\rm CMB}$ leaves the comoving Hubble radius until the end of inflation, which we denote as $N_k$. We can compute $N_k$ by tracing back the expansion history of the universe \cite{Liddle:2003as}. For this purpose, let us assume that after the perturbative decay of the inflaton, the universe stays radiation-dominated until BBN, and then follows the standard expansion history, i.e.~the universe remains radiation-dominated until the matter-radiation equality, then stays matter-dominated until entering the current dark energy era. For the pivot scale $k_{\rm CMB}=0.05{\rm Mpc}^{-1}$, the value of $N_k$ is then given by the following expression (see e.g.~\cite{Planck:2018jri}),
\be N_k \simeq 61.5+\frac{1}{4}\ln \frac{V_k^2}{\mpl^4\rho_{\rm end}} - \frac{1-3\bar{w}_{\rm rd}}{4}N_{\rm rd} -\frac{1}{12}\ln g_{\rm th}  \ , \label{eq:Nk} \ee
where $\rho_{\rm end}$ denotes the total energy density at the end of inflation, $g_{\rm th}$ the relativistic degrees of freedom at radiation domination, and $V_k \equiv V (\phi_k)$ the value of the inflaton potential when the pivot scale $k_{\rm CMB}$ leaves the Hubble radius (the value of $\phi_k$ is given in Eq.~\eqref{eq:phik} of App.~\ref{App:InflatonModel}). Following \cite{Planck:2018jri}, we fix the relativistic degrees of freedom to $g_{\rm th} =10^3$. The values of $N_{\rm rd}$ and $\bar{w}_{\rm rd}$ are then the only unknowns in Eq.~\eqref{eq:Nk}, which we have now determined with our simulations.\footnote{Note that $V_k$ itself depends on $N_k$. Therefore, we have solved Eq.~\eqref{eq:Nk} iteratively.}

\begin{figure}
    \includegraphics[width=0.47\textwidth]{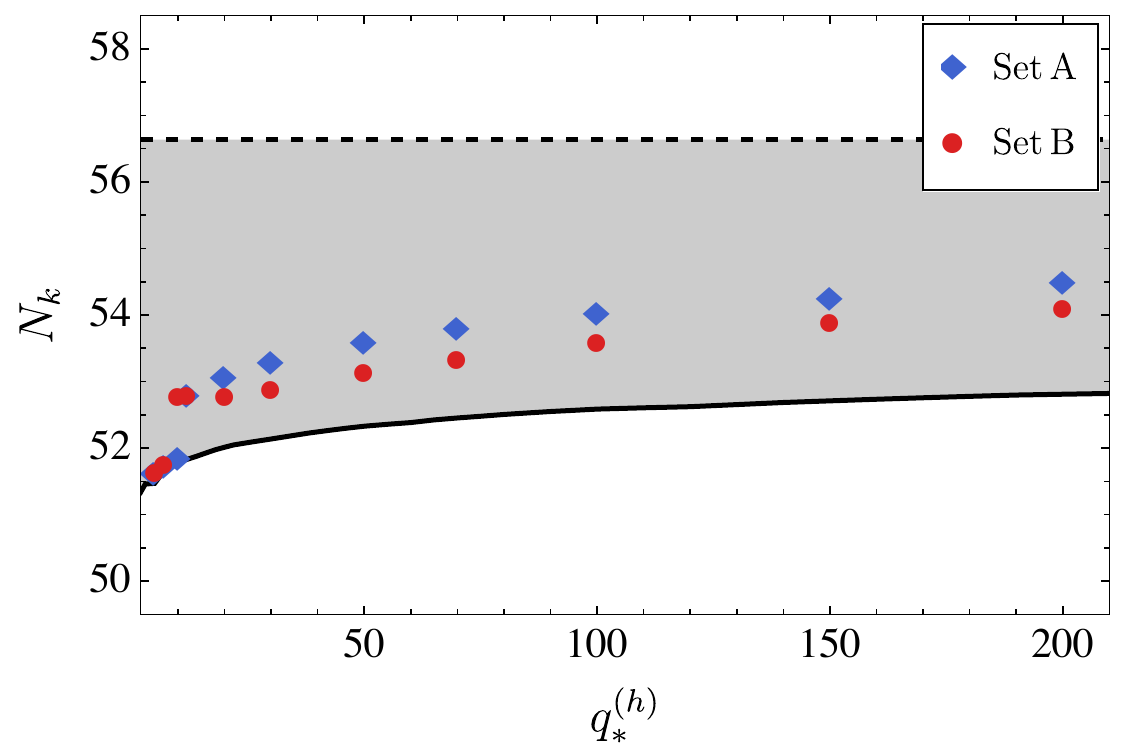}
    \caption{Number of e-folds from the moment the pivot scale $k_{\rm CMB}=0.05\rm Mpc^{-1}$ crosses the horizon until the end of inflation $N_k$ for the considered coupling parameters. The gray area depicts the theoretical uncertainty when minimal assumptions are made: the solid black line assumes a reheating phase that ignores any effects from preheating, while the dashed line assumes instant reheating, $N_{\rm rd}=0$.}
    \label{fig:Nk}
\end{figure} 

In Fig.~\ref{fig:Nk} we show the values of $N_k$ for our considered coupling parameters. Again, the gray area indicates the theoretical uncertainty of $N_k$ mentioned in Sect.~\ref{sec:RehCMB}. As expected, the values of $N_k$ for $q_*^{(h)}<q^{(h)}_{\rm min}$ agree with the lower bound for $N_k$, while for $q_*^{(h)}>q^{(h)}_{\rm min}$, they are $\delta N_k \approx 0.7-1.5$ larger. \newline

\begin{figure}
    \includegraphics[width=0.47\textwidth]{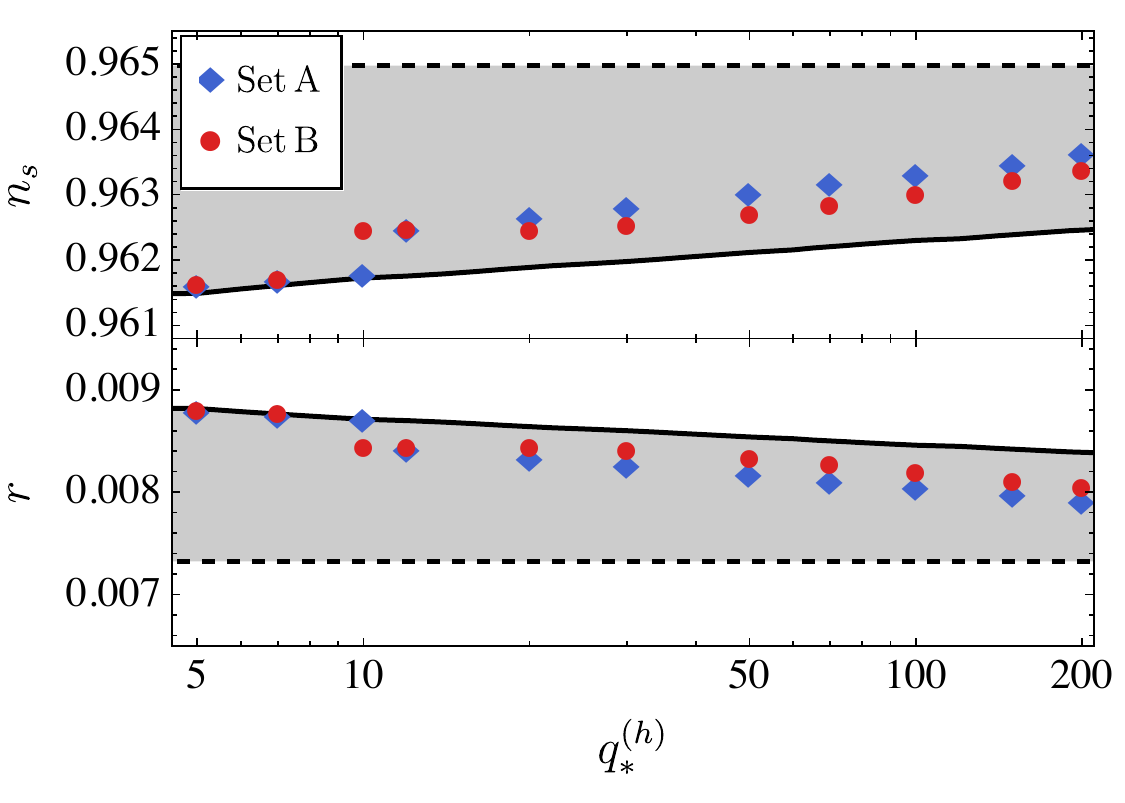}
    \caption{Predictions for the spectral index $n_s$ (top) and tensor-to-scalar $r$ (bottom) as a function of $q_*^{(h)}$ obtained from several simulations from Set A (blue) and B (red). The gray area depicts the theoretical uncertainty when minimal assumptions are made: the solid black line assumes a reheating phase that ignores any effects from preheating, while the dashed line assumes instant reheating, i.e.~$N_{\rm rd}=0$.}
    \label{fig:ns_and_r}
\end{figure}

{\bf Predictions of CMB observables:} With the obtained values of $N_k$, we can now provide more accurate predictions of the CMB observables: the spectral tilt $n_s$ and the tensor-to-scalar ratio $r$. We provide their expressions for the $\alpha$-attractor model as a function of $N_k$ in Eqs.~\eqref{eq:nsT} and \eqref{eq:tensor-to-scalarT} of App.~\ref{App:InflatonModel}. In Fig.~\ref{fig:ns_and_r} these predictions are shown for the two sets of simulations. In general, larger values of $q^{(h)}_*$ yield larger values of $n_s$, but smaller values of $r$. We can clearly observe that as soon as the non-linear dynamics of the preheating stage becomes relevant, the results for $n_s$ are shifted by $\delta n_s \lesssim 10^{-3}$ towards larger values and for $r$ by $\delta r \lesssim 5 \cdot 10^{-4}$ towards smaller values. This shows the relevance of accurately characterizing the post-inflationary evolution of the equation of state in order to provide accurate predictions of the CMB observables. 

\section{Gravitational wave signatures} \label{sec:GWsignatures}

Preheating constitutes a strong source of gravitational waves from the early universe \cite{Easther:2006gt,GarciaBellido:2007dg,GarciaBellido:2007af,Dufaux:2007pt,Dufaux:2008dn,Dufaux:2010cf,Bethke:2013aba,Bethke:2013vca,Figueroa:2017vfa,Adshead:2018doq,Adshead:2019lbr,Adshead:2019igv}. By using e.g.~3+1-dimensional lattice simulations, one can typically compute an accurate prediction for the corresponding GW spectrum at the time $u_{\rm f}$ when the GW production stops being efficient \cite{Figueroa:2011ye,GWTechNote},
\be \Omega_{\rm gw}^{\rm (f)} \equiv \frac{1}{\rho_c} \frac{d \rho_{\rm gw}}{d \log k} (k,u_{\rm f}) \ , \ee
where we have normalized the spectrum by the critical energy density $\rho_c$ as customary. However, in order to accurately predict the shape of the GW spectrum today, one needs to redshift its amplitude and frequency from the end of GW production (typically shortly before the end of the 3+1-dimensional simulation) until the present era. One can show that the amplitude and frequency redshift as follows (see App.~\ref{App:GWDilution} for a demonstration),
\begin{align}
f_{\rm gw} &\simeq 4 \cdot 10^{10} \epsilon_{\rm f}^{1/4} \frac{k}{a_{\rm f} H_{\rm f}} \left( \frac{H_{\rm f}}{m_p} \right)^{1/2} {\rm Hz} \ , \\
h_0^2 \Omega_{\rm gw} &\simeq 1.6 \cdot 10^{-5} \, \epsilon_{\rm f} \, \Omega_{\rm gw}^{\rm (f)} \ , 
\end{align}
where $a_{\rm f}$ and $H_{\rm f}$ are the scale factor and Hubble parameters at time $u = u_{\rm f}$, and $\epsilon_{\rm f}$ is the following ``suppression factor'',
\be \epsilon_{\rm f} \equiv \left( \frac{a_{\rm f}}{a_{\rm rd}}\right)^{1 - 3 \bar{w}_{\rm f}} \ , \ee
where $\bar{w}_{\rm f}$ is the average equation of state from time $u_{\rm f}$ until the achievement of radiation domination. This factor accounts for the unknown expansion rate during the reheating period: we have $\epsilon_{\rm f}= 1$ if the universe remains exactly radiation-dominated during reheating, while $\epsilon_{\rm f} < 1$ if e.g.~an intermediate period of matter-domination exists. In the second case, the amplitude of the GW spectrum may get severely suppressed and their frequencies shifted to the IR.

For our present purposes, it is convenient to express the above formulas in terms of the following two factors,
\be \epsilon_{\rm i} \equiv \left( \frac{a_{\rm end} }{a_{\rm f}}\right)^{1 - 3 \bar{w}_{\rm i}}  \ , \,\,\,\,\,\,\,
\epsilon_{\rm rd} \equiv \left( \frac{a_{\rm end}}{a_{\rm rd}}\right)^{1 - 3 \bar{w}_{\rm rd}}      \ , \ee
where $\epsilon_{\rm i}$ accounts for the expansion from the end of inflation until the end of gravitational wave production (with $\bar{w}_{\rm i}$ the average equation of state in that period) and $\epsilon_{\rm rd}$ for the expansion from the end of inflation until the onset of radiation domination (with $\bar{w}_{\rm rd}$ the average equation of state in that period). One can show that $\epsilon_{\rm f}\equiv \epsilon_{\rm i}^{-1} \epsilon_{\rm rd}$ by propagating the GW spectrum back to the end of inflation and then forward to the onset of radiation domination (see App.~\ref{App:GWDilution} for a derivation). The production of the GW peak from preheating usually happens within the first few e-folds after the end of inflation, so $\epsilon_{\rm i} \sim 0.1$, while $\epsilon_{\rm rd}$ can be fully determined with our 2+1 lattice simulations.

\begin{figure}
    \includegraphics[width=0.47\textwidth]{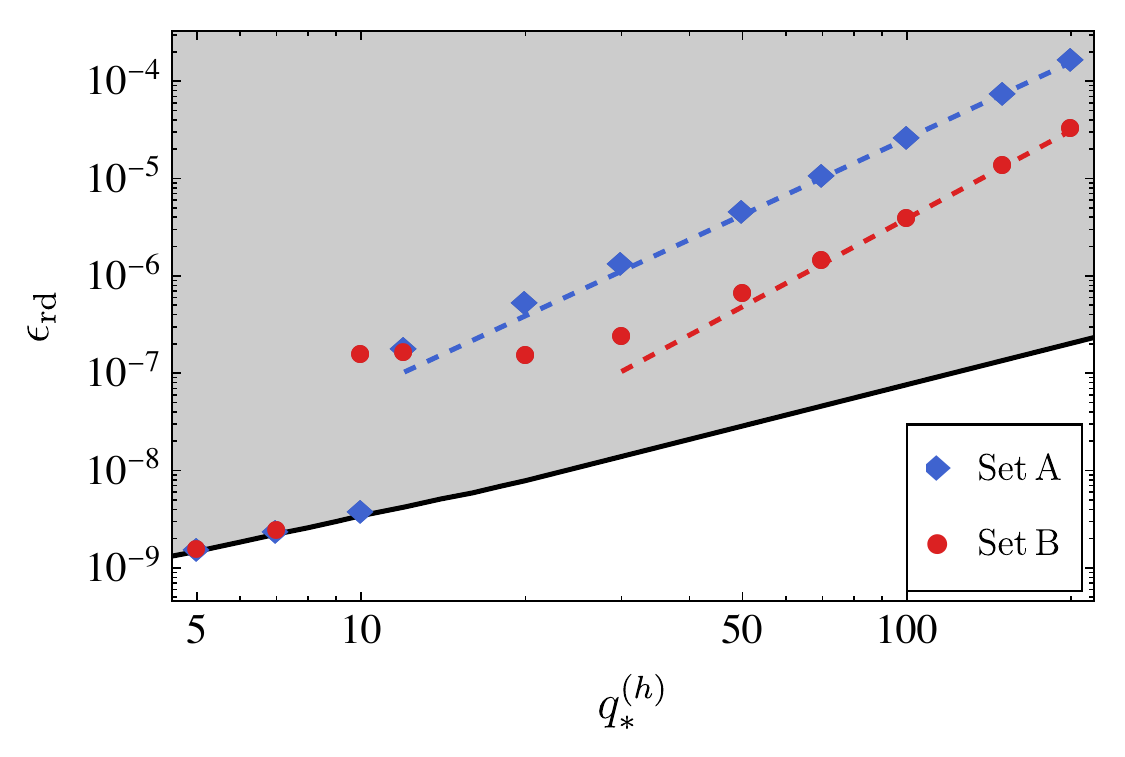}
    \caption{GW suppression factor $\epsilon_{\rm rd}$ for the considered coupling parameters. The black line shows a matter-dominated reheating phase, while the upper bound  $\epsilon_{\rm rd}=1$ would correspond to an instant reheating scenario with $N_{\rm rd}=0$. The dashed lines show the fits \eqref{eq:fitEpsilonI1} and \eqref{eq:fitEpsilonI2}.}
    \label{fig:epsilon_i}
\end{figure}

In Fig.~\ref{fig:epsilon_i} the values of $\epsilon_{\rm rd}$ for the considered coupling parameters. The black line shows the case of a reheating phase that ignores the preheating dynamics. The dashed lines indicate the following two fits, which we have found for all $ q_*^{(h)}> q_{\rm min}^{(h)}$,
\begin{align} 
\text{Set A:}\,\,\,\, \epsilon_{\rm rd} &\simeq  (8.4 \pm 0.1) \cdot 10^{-8} \left( \frac{ q_*^{(h)} }{10} \right)^{2.60 \pm 0.01 } \ ,\label{eq:fitEpsilonI1} \\
\text{Set B:}\,\,\,\, \epsilon_{\rm rd} &\simeq (3.8 \pm 0.3) \cdot 10^{-9} \left( \frac{ q_*^{(h)} }{10} \right)^{3.01 \pm 0.03 }  \ . 
\label{eq:fitEpsilonI2}
\end{align}
Therefore, the post-inflationary expansion history leads to a severe suppression of the GW spectrum amplitude from preheating of several orders of magnitude for the range of coupling constants considered in this work, as well as to a shift of the relevant frequencies towards the infrared\footnote{Note that extrapolating these formulas to very large couplings would lead to $\epsilon_{\rm rd} \gg 1$, which is not possible in our set-up. In this regime, we expect these scaling laws to change in such a way that $\epsilon_{\rm rd} \rightarrow 1$ in the limit $q_*^{(h)} \rightarrow \infty$.}.

In order to obtain a more accurate prediction for the expected gravitational wave signal from preheating with trilinear interactions, we can combine our results for $\epsilon_{\rm rd}$ with the results from \cite{Cosme:2022htl}, which studied the GW production in the same set-up with 3+1-dimensional lattice simulations. There, they obtained the following parameterizations for the frequency and amplitude of the main spectral peak as a function of $q_*^{(h)}$,
\begin{align}
f_{\rm gw}  &\simeq (1.0 \pm 0.1) \cdot 10^8\, \epsilon_{\rm f}^{1/4} \left( \frac{q_*^{(h)}}{10} \right)^{0.52 \pm 0.04} {\rm Hz} \ , \nonumber \\    
h_0^2 \Omega_{\rm gw}^{(0)} &\simeq (2.67 \pm 0.5) \cdot 10^{-9}\, \epsilon_{\rm f} \left( \frac{q_*^{(h)}}{10} \right)^{-0.43 \pm 0.07} \ , \label{eq:latticeGWparams} 
\end{align}
valid for Set A and $q_*^{(h)} > 10$. Their simulations extend until time $u_{\rm f} = 400$. With such a relatively short simulation time one might presume that the universe continues to evolve with an approximately radiation dominated equation of state, and set $\epsilon_{\rm f} = 1$, which yields frequencies of $\sim 10^{8} {\rm Hz}$ and amplitudes of $h_0^2 \Omega_{\rm gw}^{(0)} \gtrsim 10^{-9}$ for the range of couplings considered in this work. However, as we have shown, this is not correct. We can now use our results for the equation of state, which takes this effect into account, in order to obtain more accurate predictions for the GW spectrum today. By combining \eqref{eq:latticeGWparams} with \eqref{eq:fitEpsilonI1} and $\epsilon_{\rm i}\approx0.076(q_*^{(h)}/10)^{0.12}$, we obtain the following predictions for the frequency and amplitude of the main peak,
\begin{align}
f_{\rm gw} &\simeq (3.2 \pm 0.4) \cdot 10^6  \left( \frac{q_*^{(h)}}{10} \right)^{1.14 \pm 0.04} {\rm Hz} \ , \nonumber \\    
h_0^2 \Omega_{\rm gw}^{(0)}  &\simeq (2.8 \pm 0.9) \cdot 10^{-15} \left( \frac{q_*^{(h)}}{10} \right)^{2.05 \pm 0.09} \hspace{-0.3cm} \ . 
\end{align}
Therefore, for the example of parameter Set A and for $q_*^{(h)} > 10$, the post-inflationary expansion history leads to a suppression of up to six orders of magnitude for the weaker couplings compared to the estimate with $\epsilon_{\rm f} = 1$, as well as shifts the frequency one or two orders of magnitude towards the infrared. Moreover, the dependence of the suppression factor on $q_*^{(h)}$ leads to a stronger dependence of the position and amplitude of the peak on the coupling strength, and in fact, the amplitude of the signal now increases with $q_*^{(h)}$. While the gravitational wave signal is in any case well outside the range of detection of planned GW observatories, our study shows the importance of correctly characterizing the post-inflationary expansion rate in order to obtain accurate predictions. This fact could be particularly relevant in light of future proposals for ultra-high frequency GW detectors \cite{Aggarwal:2020olq,Aggarwal:2025noe}.

\section{Summary and discussion} \label{sec:summary} 

The post-inflationary expansion history of the universe can greatly affect the observational implications of a given inflationary model. Therefore, it is essential to fully determine it as accurately as possible, including also the impact of any possible non-perturbative process that might occur in this period, such as preheating. In this work, we have focused on a paradigmatic scenario consisting of an inflaton with a quadratic potential, coupled to a (effectively massless) daughter field through a trilinear interaction, which also has a quartic self-interaction for stability purposes. By combining the results from both 2+1-dimensional lattice simulations and Boltzmann equations, we have been able to completely characterize the post-inflationary expansion history from the end of inflation until the completion of perturbative reheating, and leverage such information to obtain precise predictions for two observational windows: the inflationary CMB observables $n_s$ and $r$, and the stochastic background of gravitational waves from preheating measured today.

For sufficiently strong couplings, the daughter field experiences tachyonic resonance, which fragments the inflaton and deviates the equation of state from $\bar{w} = 0$ towards a maximum value $\bar{w} \rightarrow \bar{w}_{\rm max}$. However, the fragmentation is only partial and the universe never becomes completely radiation-dominated. Moreover, our results exhibit that such fragmentation is only temporary: at later times the interactions stop exciting the fields efficiently and their fluctuations start diluting as radiation. Consequently, the homogeneous mode of the inflation becomes again the dominant energy density source, pushing the equation of state towards $\bar{w} = 0$. To our knowledge, our simulations are the first ones that explicitly show the return of the effective equation of state back towards zero after the completion of preheating in the presence of trilinear interactions with a massless daughter field. In our model, the transition towards a radiation-dominated universe happens instead at much later times, approximately $\sim 15-22$ e-folds after the end of inflation for the considered coupling constants, when the inflaton decays perturbatively. This evolution of the equation of state is qualitatively similar for a diverse range of coupling strengths, as well as when quadratic-quadratic interactions between both fields also exist, see App.~\ref{App:Quadratic}.

With our results we have determined the number of e-folds $N_{\rm rd}$ from the end of inflation until the definitive achievement of radiation domination, as well as the average equation of state $\bar{w}_{\rm rd}$ during that period. Our predictions take into account the preheating dynamics, which strongly affect the values of $N_{\rm rd}$ and $\bar{w}_{\rm rd}$, and have been obtained for different choices of coupling constants. With these, we have been able to narrow down more precisely the value of $N_k$ (i.e.~the number of e-folds of expansion when the pivot scale leaves the horizon until the end of inflation). In the considered examples, the existence of preheating changes the value of $N_k$ by more than one e-fold, and in fact, this deviation is larger than the one found for $\sim \phi^2X^2$ couplings, for which the change is always $\delta N_k<1$ \cite{Antusch:2021aiw}. This shifts the value of $n_s$ towards larger values, $\delta n_s \lesssim 0.001$, and $r$ towards smaller values, $\delta r \lesssim 5 \cdot 10^{-4}$, see Fig.~\ref{fig:ns_and_r}, with the shifts being larger for larger coupling parameters. This result may become relevant in the light of future CMB experiments \cite{LiteBIRD:2022cnt}. In fact, one could potentially use this information to observationally constrain the coupling strengths between the fields \cite{Drewes:2017fmn}. On the other hand, the long period of almost matter-dominated expansion significantly affects the frequencies and amplitudes at which the stochastic GW background generated during preheating is observed today: the amplitude gets suppressed between two and six orders of magnitude for the range of couplings considered here, while the frequency gets shifted between one and two orders of magnitude to the infrared. This effect is stronger for smaller couplings, which leads to a different dependence of the frequency and amplitude of the main spectral peak on the coupling parameter: the frequency now scales as $f_{\rm gw} \sim (q_*^{(h)})^{1.1}$ and the amplitude as $\Omega_{\rm gw} \sim (q_*^{(h)})^{2.1}$.

Although we have focused on single-field inflationary potentials with a quadratic minimum in this paper, the techniques developed here to fully characterize the post-inflationary equation of state could be applied in various other models of interest. More specifically, one could simulate the initial preheating dynamics of a given model with lattice simulations, in order to fully capture the non-linearities of the equations of motion, and once preheating ends, switch to an effective picture, in which the corresponding Boltzmann equations are solved until the achievement of radiation domination. This way, one could obtain exact predictions for the inflationary CMB observables and the stochastic gravitational wave background measured today for those models. Moreover, lattice simulations in 2+1 dimensions can be an efficient way of realistically simulate the preheating dynamics without the computational overload of three-dimensional simulations (although the suitability of 2+1-dimensional simulations to mimic three-dimensional dynamics must be proven on a case by case basis, as we have done in App.~\ref{App:3D}).

In our work, we have restricted ourselves to inflaton potentials that can be approximated by a quadratic function during preheating, but it would be interesting to extend our work to other kinds of scenarios. For example, one could consider an inflaton oscillating around regions flatter than quadratic, like in the $\alpha$-attractor T-model case \eqref{eq:alphaattractor} for $M \lesssim \mpl$. In the absence of external couplings, the inflaton is expected to fragment into oscillons \cite{Amin:2011hj}. Oscillon formation can also occur in the presence of daughter field interactions \cite{Antusch:2015ziz,Shafi:2024jig}, with the oscillon lifetime depending on the coupling type and strength. It would be interesting to study the evolution of the equation of state of the universe in such scenarios. It would also be interesting to consider inflaton potentials behaving as $V(\phi) \propto |\phi|^p$ with $p \geq 2$ around the minimum: in this case, we expect the equation of state to remain radiation-dominated after preheating because the inflaton experiences a maintained self-resonant growth of its own fluctuations \cite{Lozanov:2016hid,Lozanov:2017hjm}. This would in particular alleviate the suppression of the gravitational wave spectrum from preheating. Similarly, one could consider scenarios in which the daughter field is coupled sufficiently strongly to a fermion. 
This introduces a perturbative decay channel that can enhance the depletion of energy from the inflaton, potentially extending the nearly radiation-dominated phase \cite{Garcia-Bellido:2008ycs,Bezrukov:2008ut,Repond:2016sol,Fan:2021otj,Mansfield:2023sqp}. Finally, another interesting extension could consist in incorporating gravitational effects into the analysis, which can fragment an inflaton with a quadratic potential even in the absence of interactions to other fields some e-folds after the end of inflation \cite{Musoke:2019ima}. These effects could potentially affect the evolution of the equation of state. 

\begingroup
\renewcommand{\addcontentsline}[3]{}
\acknowledgments
K.M. thanks D.~G.~Figueroa and N.~Loayza for useful discussions. FT (ORCID 0000-0003-1883-8365) is supported by a \textit{Beatriu de Pinós} fellowship (reference number:~2022 BP 00063) from the Ministry of Research and Universities of the Government of Catalonia. F.T. also acknowledges financial support from grants PID2022-137268NB-C52 from the Spanish Ministry of Science and Innovation, 2021-SGR-00872 from AGAUR, and CEX2024-001451-M funded by MICIU/AEI/10.13039/501100011033. KM~acknowledges support from the Swiss National Science Foundation (project number P500PT\_214466)

\appendix

\section{Tachyonic resonance in the presence of quadratic-quadratic interactions}\label{App:Quadratic}

In this Appendix we discuss the post-inflationary dynamics of models in which the inflaton is coupled to a daughter field through both trilinear and quadratic-quadratic interactions. The corresponding potential is given by,
\be
V (\phi,X) = V_{\rm inf} (\phi) + \frac{1}{2}h\phi X^2 + \frac{1}{2} g^2 \phi^2 X^2 + \frac{1}{4}\lambda X^4 \ , \label{eq:model2}
\ee
where $g$ is a dimensionless coupling constant. This potential gives rise to an excitation of daughter field modes through both \textit{tachyonic} and \textit{parametric} resonance \cite{Kofman:1994rk}. We discuss the interplay of these processes in the following.

Analogously to Eqs.~\eqref{eq:respar} and \eqref{eq:resself} we define a resonance parameter associated to the quadratic-quadratic interaction as follows,
\be
\tilde{q}^{(g)} \equiv  q_*^{(g)} a^{-3} \ , \hspace{0.8cm} q_*^{(g)} \equiv \frac{g^2 \phi_*^2}{\omega_*^2} \ .
\ee
By following the same rescaling and linearization procedure of Sect.~\ref{sec:LinearAn}, we obtain the following linearized equations for the homogeneous mode of the inflaton and the daughter field modes in the Hartree-Fock approximation,
\begin{align}
\bar{\varphi}''+ (1+\tilde{q}^{(g)}\langle \chi^2\rangle )\bar{\varphi}+\frac{1}{2}\tilde{q}^{(h)}\langle \chi^2\rangle &= 0 \ , \label{eq:eomvarphi2}\\
\delta \chi^{''}_{k} + (\tilde{\kappa}^2 + \tilde{m}_{\chi, {\rm eff}}^2)\delta \chi_{k}&= 0 \label{eq:modechi2} \ , 
\end{align}
where the effective mass of the daughter field in natural units $\tilde{m}_{\chi} \equiv m_{\chi} / \omega_*$ is now,
\be
\tilde{m}_{\chi, {\rm eff}}^2 \equiv  \tilde{q}^{(h)} \bar{\varphi} + \tilde{q}^{(g)} \bar{\varphi}^2 + 3 \tilde{q}^{(\lambda)}\langle \chi^2\rangle \label{eq:m_effB} \ .
\ee
Let us introduce the following time-dependent ratio between the effective resonance parameters of the trilinear and quadratic interactions,
\be \tilde{R}\equiv \frac{\tilde{q}^{(g)}}{\tilde{q}^{(h)}} = R_*a^{-3/2} \ , \hspace{0.5cm} R_* \equiv \frac{q_*^{(g)}}{q_*^{(h)}} \ ,
\ee
where $R_*$ is the value of the ratio at the end of inflation. This function represents the relative strength of parametric vs.~tachyonic resonance. This ratio decreases over time, i.e.~as the universe expands the trilinear interaction becomes increasingly more dominant. We can hence identify two limiting regimes:

    \begin{figure*}
        \centering
        \includegraphics[width=0.48\textwidth]{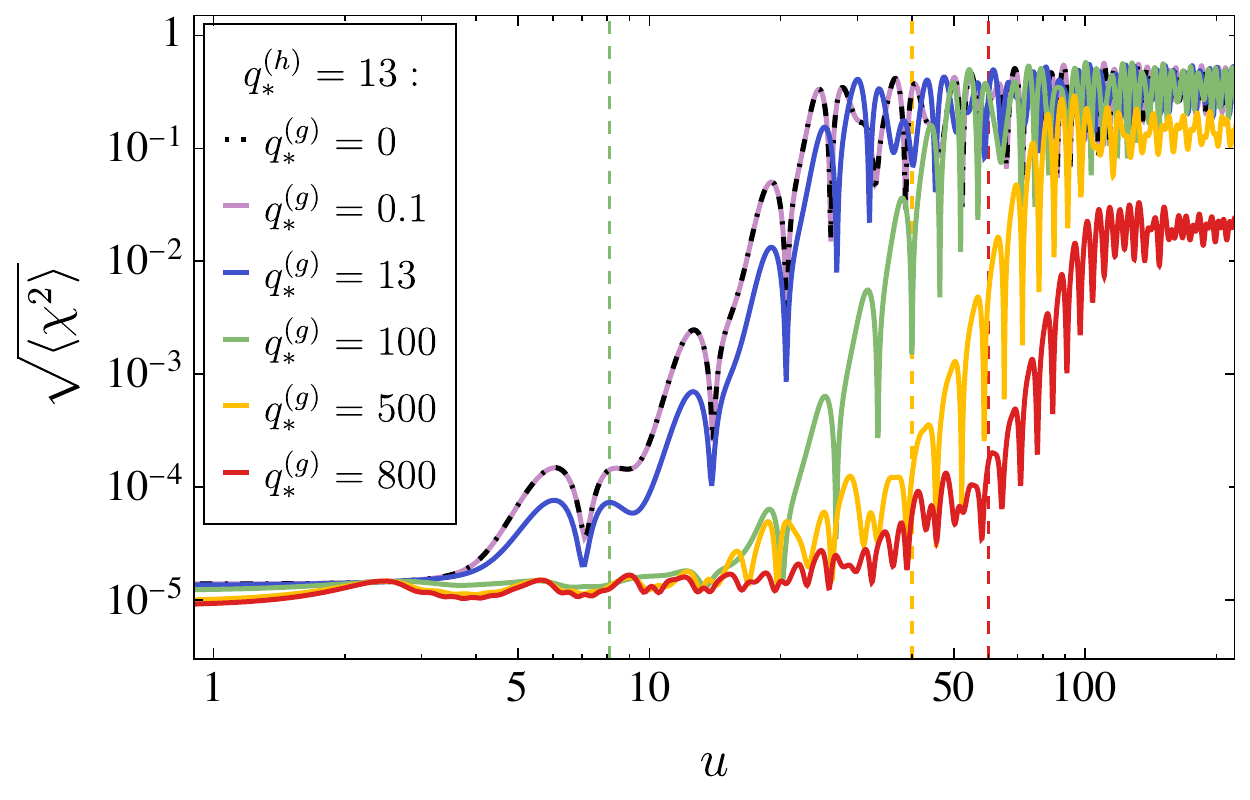} \hspace{1em}
    \includegraphics[width=0.48\textwidth]{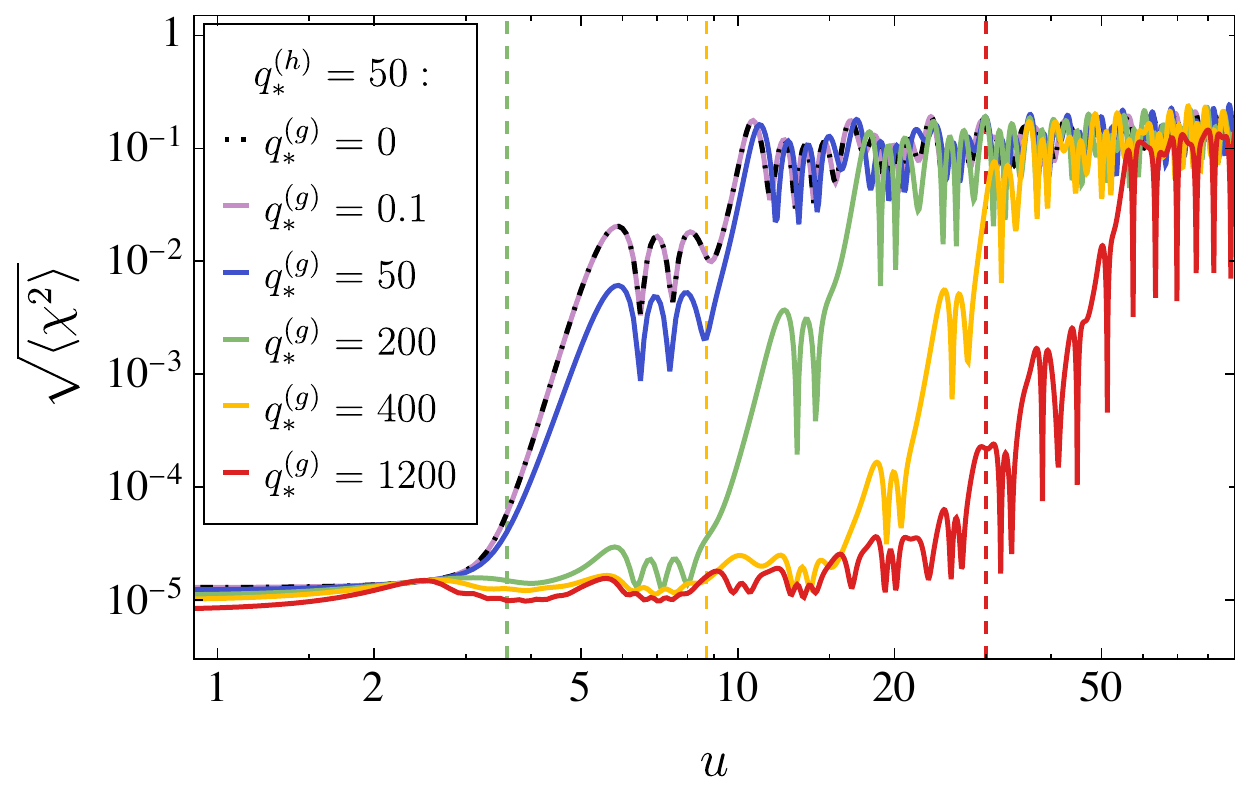}
        \caption{\textit{Left:} Evolution of $\sqrt{\langle \chi^2 \rangle}$ for $q_*^{(h)}=13,\,50$ (left and right panels respectively), $q_*^{(\lambda)}=(q_*^{(h)})^2$, and different values of $q_*^{(g)}$. The evolution is obtained by solving Eqs.~\eqref{eq:eomvarphi2}-\eqref{eq:modechi2} self-consistently with the Friedmann equations. The dashed vertical lines indicate when $\tilde{R}\geq1$ for each case, with their colours matching the ones of the respective cases.}
        \label{fig:VarTRvsPR}
\end{figure*}

\begin{itemize}[left=0pt, labelindent=3pt] 
\item[{\bf a)}] $\pmb{ R_*<1}${\bf :} In this case, tachyonic resonance dominates over parametric resonance throughout the entire linear preheating stage, so the quadratic-quadratic interactions can be ignored. Hence, the energy ratios and equation of state evolve as discussed in the main text. This is well illustrated in Fig.~\ref{fig:VarTRvsPR}, which shows that the evolution of the daughter field variance $\sqrt{\langle \chi^2\rangle}$ under the Hartree-Fock approximation is identical for both $q_*^{(g)}=0$ and $q_*^{(g)}=0.1 \ll q_*^{(h)}$.

\item[{\bf b)}] $\pmb{ R_*> R_{\rm crit} \approx 200}$: In this limit, the resonant effect stemming from the quadratic-quadratic interaction dominates over the one of the trilinear interaction, so that the daughter field never becomes tachyonic. The effective mass  $\tilde{m}_{\chi, \rm eff}$ \eqref{eq:m_effB} is then,
\be
\tilde{m}_{\chi, {\rm eff}}^2 \simeq  \tilde{q}^{(g)} \bar{\varphi}^2 + 3 \tilde{q}^{(\lambda)}\langle \chi^2\rangle \ .
\ee
In this case, the resonant amplification of daughter field modes and later evolution of the equation of state is driven by parametric resonance, which has been studied in detail in \cite{Antusch:2020iyq,Antusch:2021aiw} (for the case in which the daughter field does not have a quartic self-interactions) and \cite{Antusch:2022mqv} (when such self-interaction exists). From the results of those works, we gather the following conclusions: 

\begin{enumerate}
    \item For $q_*^{(g)}<q_{\rm min}^{(g)}\approx 10^3$, the system enters the narrow resonance regime before the daughter field fluctuations fragment the inflaton. 
    \item For $q_*^{(g)} < q_*^{(\lambda)}$, the quartic self-coupling leads to an early termination of the resonance, and the variance saturates at $\langle \chi^2\rangle\approx q_*^{(g)}\bar{\varphi}^2/(3q_*^{(\lambda)})$.
    \item For $q_*^{(g)} > q_*^{(\lambda)}$ the fluctuations of the daughter field grow exponentially until backreaction processes become relevant, fragmenting the inflaton. In the latter case, this happens approximately when the condition $\rho_{\bar \phi}\simeq \rho_X$  holds, at time $u_{\rm br}\sim10^2$, and the saturated variance of the daughter field scales as $\langle \chi^2\rangle\sim (q_*^{(g)})^{-3/2}$ \cite{Khlebnikov:1996zt}.
\end{enumerate}

In the latter case, the exact details on the resonant growth may differ when a small trilinear interaction is also present, but by solving numerically the equations of motion in the Hartree-Fock approximation, we have observed that the value of $u_{\rm br}$ coincides with the one mentioned above for $R_* \geq R_{\rm crit} \approx 200$.

\end{itemize}

On the other hand, more intricate evolutions appear in the intermediate regime $1 \lesssim R_* \lesssim R_{\rm crit}$. In these cases, the daughter field experiences first a short phase of (relatively weak) parametric resonance induced by the quadratic-quadratic interaction, while the tachyonic resonance only becomes dominant once $\tilde{R}\leq 1$. In other words, the existence of the quadratic-quadratic interaction leads to a delay of the tachyonic growth of the daughter field modes, which can even lead to its complete suppression if the condition $\tilde{q}^{(h)}< q_{\rm min}^{(h)}$ occurs before $\tilde{R}\leq 1$. We can clearly see this in Fig.~\ref{fig:VarTRvsPR}, where we show the evolution of $\sqrt{\langle \chi^2\rangle}$ for several cases in this regime, for $q_*^{(h)}=13,\,50$ (left and right panels respectively), $q_*^{(\lambda)}=(q_*^{(h)})^2$, and different choices of $q_*^{(g)}$: we observe that the tachyonic growth of  $\sqrt{\langle \chi^2\rangle}$ only starts when $\tilde{R}=1$ (indicated for each case by vertical dashed lines). And in the case of $q_*^{(g)}=800$ in the left panel, $\sqrt{\langle \chi^2\rangle}$ stops growing before reaching its expected value.\\

{\bf Evolution of equation of state:} To conclude, here we present results for the equation of state when the inflaton is coupled to a daughter field through both trilinear and quadratic-quadratic interactions, obtained with 2+1-dimensional lattice simulations. As in the bulk text, we simulate model (\ref{eq:model2}) with the $\alpha$-attractor T-model potential (\ref{eq:alphaattractor}), for $M=5\mpl$ and $N_k=55$. In Fig.~\ref{fig:tr_vs_pr_lat} we depict the evolution of the equation of state $w$ (blue) and its oscillation average $\bar{w}$ (red) for two example cases with $q_*^{(h)} >0$ and $q_*^{(g)}>0$. For comparative purposes, we also show the evolution of the effective equation of state when either the trilinear or the quadratic-quadratic interaction is shut down, i.e.~$q_*^{(g)}=0$ or $q_*^{(h)} = 0$. For all cases we have fixed $q_*^{(\lambda)}=(q_*^{(h)})^2$.

\begin{figure*}
    \centering
        \includegraphics[width=0.48\textwidth]{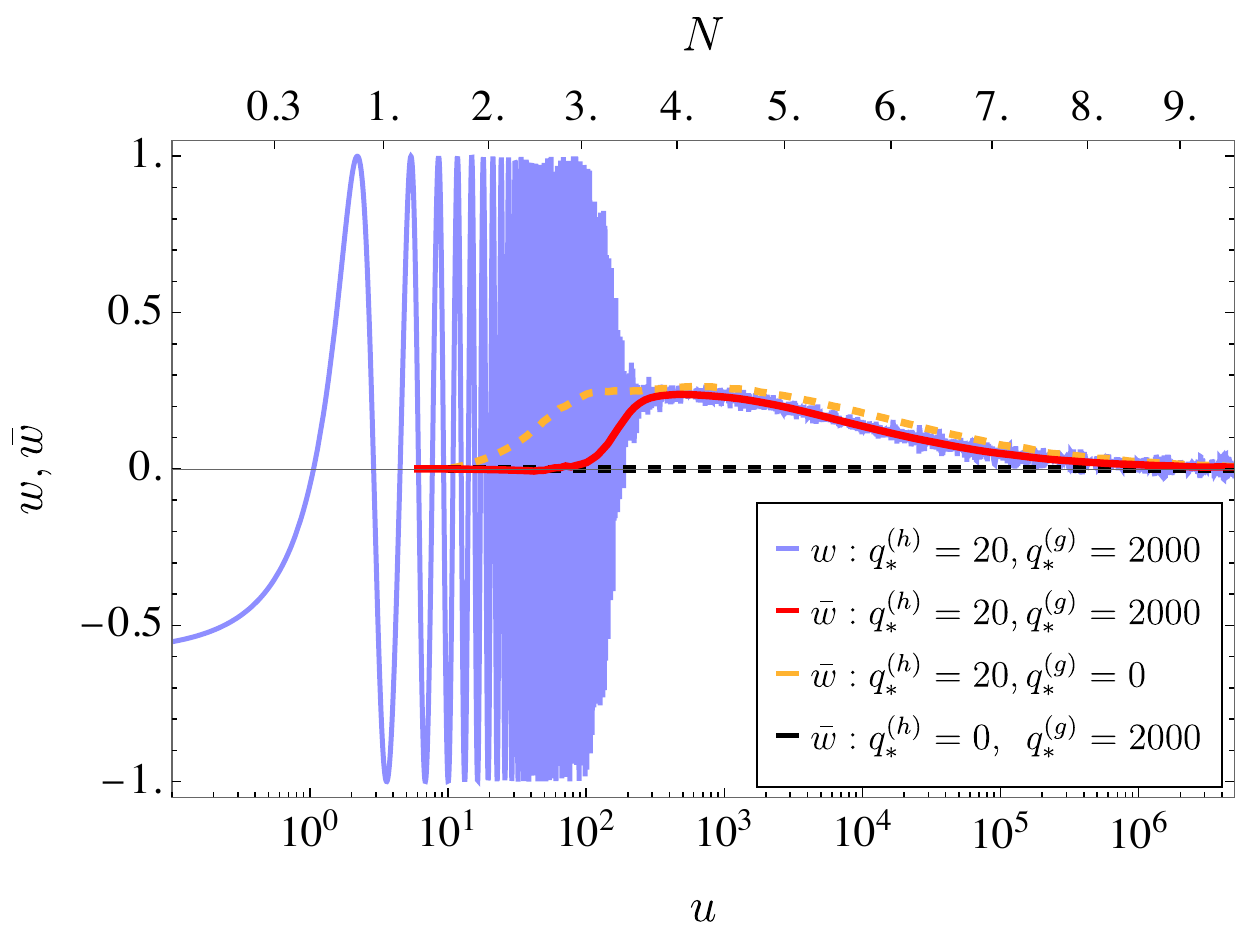} \hspace{1em}
    \includegraphics[width=0.48\textwidth]{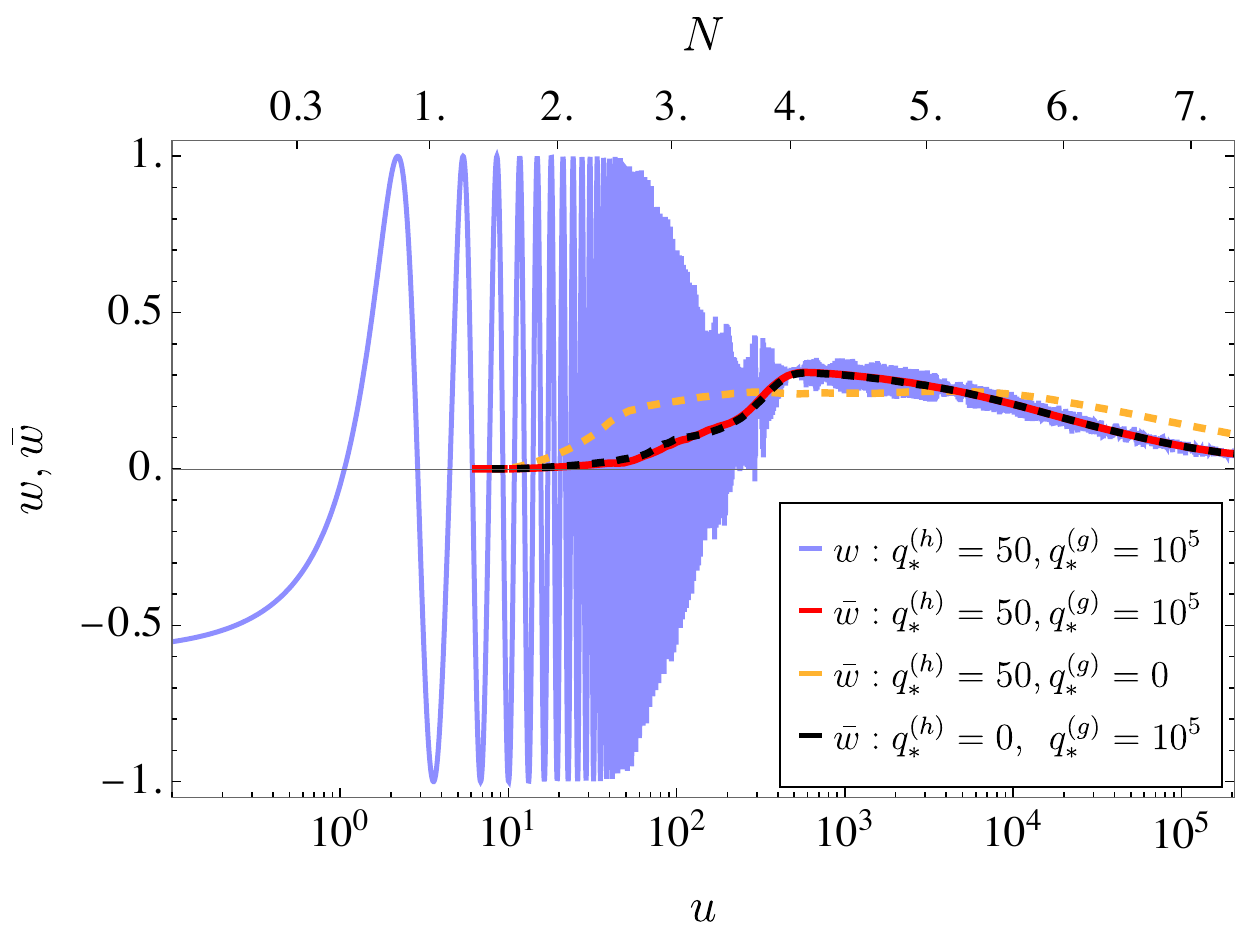}
        \caption{Evolution of equation of state $w$ (blue) and its oscillation average $\bar{w}$ (red) for $q_*^{(h)}=20$ and $q_*^{(g)}=2000$ (left panel), as well as for $q_*^{(h)}=50$ and $q_*^{(g)}=10^5$ (right panel). In both cases we have fixed the self-coupling parameter to $q_*^{(\lambda)}=(q_*^{(h)})^2$. The dashed lines indicate the evolution of the effective equation of state $\bar{w}$ for either $q_*^{(g)}=0$ (dashed orange) or $q_*^{(h)}=0$ (dashed black), with the other coupling parameters unchanged.}
        \label{fig:tr_vs_pr_lat}
\end{figure*}

The left panel of Fig.~\ref{fig:tr_vs_pr_lat} shows the case $q_*^{(g)}=2000$ and $q_*^{(h)}=20$, i.e.~$R_*<200$. In this scenario the system experiences first a short stage of parametric resonance, which enters the narrow resonance regime before $\bar{w}$ starts to deviate relevantly from its homogeneous value (see black dashed line, which always stays at $\bar{w}\approx 0$). Then, tachyonic resonance becomes the dominant process. As a result, the inflaton homogeneous stage (as discussed in Sect.~\ref{Sec:Lattice}) survives for many more oscillations in comparison to the pure trilinear case (orange dashed line). At around $u\approx 10^2$ the daughter field modes start backreacting onto the inflaton via the quadratic-quadratic coupling and the homogeneous inflaton mode starts to fragment. During this backreaction process, the system becomes highly non-linear and the average equation of state $\bar{w}$ deviates noticeably towards radiation domination. At $u\sim 500$ the coupling parameters have decreased sufficiently and no more energy is exchanged, i.e.~the energy density of the inflaton and the daughter field dilute independently as matter and radiation. As a result, the equation of state slowly decreases. Eventually, the system is dominated again by the homogeneous inflaton mode and $\bar{w}\approx0$. 

The right panel in Fig.~\ref{fig:tr_vs_pr_lat} shows the case $q_*^{(g)}=10^5$ and $q_*^{(h)}=50$, for which $R_*\gg 200$. In this case, the quadratic-quadratic coupling dominates over the trilinear one throughout the whole resonance stage. Thus, $\bar{w}$ evolves very similarly as in the case of $q_*^{(h)}=0$ (black dashed line) and is very different to the one with $q_*^{(g)}=0$, where tachyonic resonance is the dominant resonance process (dashed orange line). \newline

\section{Further technical details} \label{App:Technical}

\subsection{Inflaton potential}\label{App:InflatonModel}

The lattice simulations presented in this work have been performed for the symmetric $\alpha$-attractor T-model \eqref{eq:alphaattractor}, see Ref.~\cite{Kallosh:2013hoa},
where $\Lambda$ and $M$ have dimensions of energy. The potential is approximately quadratic around the minimum, but develops a plateau at large field amplitudes, which makes it compatible with the observed upper bound of the tensor-to-scalar ratio obtained from CMB observations. The inflection point separating both regimes is given by $\phi_{\rm i} \equiv M{\rm arcsinh}(1/\sqrt{2}) \simeq 0.66 M$. For inflaton amplitudes $\phi \lesssim \phi_{\rm i}$ the inflaton potential can be well approximated by the quadratic function \eqref{eq:mon-pot}.

The slow-roll parameters take the following form,
\bea
\varepsilon_{V} & \equiv & \frac{\mpls}{2} \left(\frac{V_{\rm T}'}{V_{\rm T}}\right)^2 = \frac{8\mpls}{M^2\
    \sinh^2\left(\frac{2|\phi|}{M}\right)}   \label{eq:SRepsilon}\:, \\
\eta_{V} & \equiv & \mpls \frac{V_T''}{V_T} = \frac{8\left(2-\rm{cosh}\left(\frac{2|\phi|}{M}\right)\right)\mpls}{M^2\sinh^2\left(\frac{2|\phi|}{M}\right)}\label{eq:SReta} \ .
\eea
Let us define $\phi_{\rm e}$ as the field amplitude when the condition $\varepsilon_V(\phi_{\rm e}) \equiv 1$ holds, corresponding approximately to the end of inflation,
\be\phi_{\rm e} = \frac{1}{2}M{\rm arcsinh} \left(   \frac{\sqrt{8}m_{\rm pl}}{M} \right)\ .\label{eq:phiend}
\ee
Inflation happens for field amplitudes $\phi \gtrsim \phi_{\rm e}$. For values $M \gtrsim 1.6 m_{\rm pl}$ we have $\phi_{\rm i}>\phi_{\rm e}$,
so inflation ends in the positive-curvature region of the potential, and we can take the quadratic function \eqref{eq:mon-pot} as a good approximation during the post-inflationary oscillatory regime. This is the scenario considered in this work.

The observed amplitude of the primordial spectrum of scalar perturbations in the CMB, $A_s \simeq {(2.1\pm 0.1) \cdot 10^{-9}}$ \cite{Planck:2018jri}, yields a constraint $\Lambda = \Lambda (M)$  between $M$ and $\Lambda$. The inflaton amplitude $\phi_k$ at which relevant fluctuations leave the horizon can be obtained from
\be
N_k=\frac{1}{\mpl}\int^{\phi_e}_{\phi_{k}}\frac{1}{\sqrt{2\varepsilon_{V}}}d\phi \ ,
\ee
where $N_k$ denotes the number of e-folds from the moment these fluctuations leave the horizon until the end of inflation. The $\alpha$-attractor model yields,
\begin{align}
&\hspace{-0.2cm}\phi_{k}=\frac{1}{2}M{\rm arccosh}\left(\frac{\mathcal{I}(N_k)}{N_k}+\mathcal{J}\right)\ , \label{eq:phik} \\
&\hspace{0.6cm} \mathcal{I}(N_k) \equiv 8N_k^2 \frac{m_{\rm pl}^2}{M^2} \ , \hspace{0.4cm} \mathcal{J} \equiv \sqrt{1+8 \frac{m_{\rm pl}^2}{M^2}} \nonumber \ .
\end{align} 
We fix the scale of these fluctuations to the CMB pivot scale $k_{\rm cmb}=0.05{\rm Mpc}^{-1}$. By using Eq.~\eqref{eq:phik}, we are able to determine the value of $\Lambda^4$ via the expression $V_T(\phi_k)=24\pi^2\varepsilon_V(\phi_k)A_s \mpl^4$,
\begin{align} 
\Lambda^4 &=\frac{6\pi^2M^2 A_s \mpl^{2}}{N_k^2}f(M,N_k) \ ,\label{eq:Lambda4} \\
& \hspace{0.5cm} f(M,N_k) \equiv \left(\frac{\mathcal{I} (N_k)+2N_k\mathcal{J}+2N_k}{2+4N_k\mathcal{J}+\mathcal{I} (N_k)}\right)^2\ .
\end{align}
In the monomial limit $M \rightarrow \infty$ we have $f=1$, while for the values $M=5\mpl$ and $N_k=55$ simulated in Sect.~\ref{Sec:Lattice} we have $f\approx0.98$. Remarkably, the mass around the minimum $m_\phi \equiv \Lambda^2 (M,N_k)/M$ is almost constant independently of the choice of $M$, see e.g.~Fig.~3 of \cite{Antusch:2021aiw}. For example, for $N_k= 55$ it is always in the range $m_{\phi}/\mpl = [0.139 A_s, 0.140A_s]$, the lower and upper limits correspond to the $M\rightarrow 0$ and $M \rightarrow \infty$ limits. 
 
At last, we provide expressions for the inflationary CMB observables, i.e.~the \textit{spectral index} $n_s$ and the \textit{tensor-to-scalar ratio} $r$:
\begin{align} 
n_s&=1-6\varepsilon_{V}(\phi_k)+2\eta_{V}(\phi_k) \ ,\\
r &=16\varepsilon_{V}(\phi_k) \ . 
\end{align}
Using the expressions (\ref{eq:SRepsilon}), (\ref{eq:SReta}) and (\ref{eq:phik}) we obtain
\begin{align} 
n_s & = 1-2\frac{1+\mathcal{J}+\mathcal{I}  (N_k)/N_k}{1+2N_k\mathcal{J}+\mathcal{I} (N_k)}\ \label{eq:nsT},\\
r& = \frac{16}{1+2N_k\mathcal{J}+\mathcal{I} (N_k)} \ \label{eq:tensor-to-scalarT}. 
\end{align}
The observational constraints on the spectral index are the ones obtained by Planck, $n_s=0.9668 \pm 0.0037$ \cite{Planck:2018jri}. This number is slightly smaller considering the most recent results from ACT \cite{ACT:2025fju}. On the other hand, the current upper bound on the tensor-to-scalar ratio, $r< 0.036$ at $95\%$ confidence level \cite{BICEP:2021xfz}, sets the constraint $M\lesssim 8.5\mpl$.

\begin{figure*}
    \includegraphics[width=0.47\textwidth]{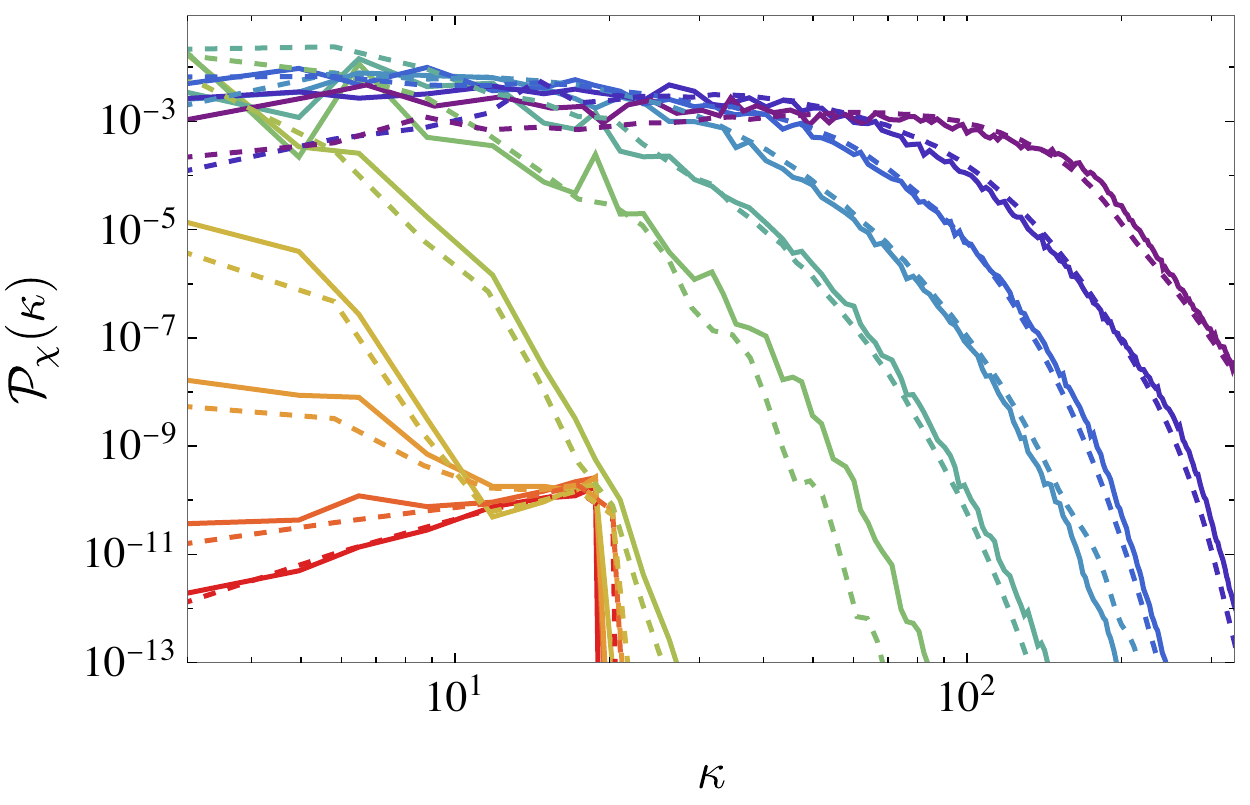}  \hspace{0.3cm}
     \includegraphics[width=0.47\textwidth]{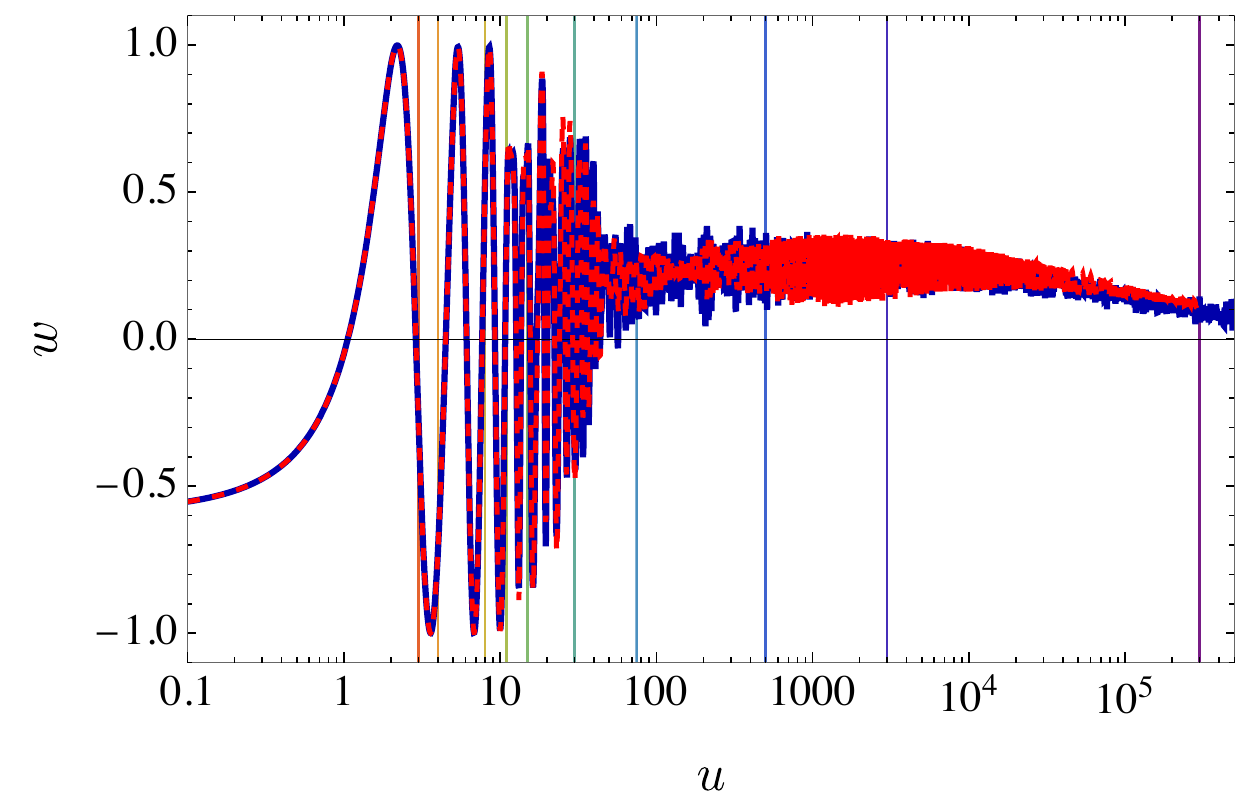}
    \caption{Left: Evolution of the daughter field spectrum obtained from simulations in 3+1 (dashed lines) and 2+1 dimensions (solid lines) for the coupling parameters $q_*^{(h)}=50$ and $q_*^{(\lambda)}=2500$. The different lines show the spectrum at different times. Right: Evolution of the equation of state $w$ in 3+1 (red dashed) and 2+1 (blue) dimensions in the same case. The vertical lines indicate the times at which the spectra in the left panel have been outputted.}
    \label{fig:Specs2D3D}
\end{figure*} 

\subsection{Energy ratios}  \label{App:EnergyEos}

The energy and pressure densities of the two-field system considered in this work are
\begin{align}
    \rho &= \frac{1}{2}\sum_f^{\phi,X}\dot{f}^2 +\frac{1}{2a^2}\sum_f^{\phi,X} |\nabla f|^2 + V(\phi, X) \ , \\
    p &= \frac{1}{2}\sum_f^{\phi,X}\dot{f}^2 - \frac{1}{6 a^2} \sum_f^{\phi,X} |\nabla f|^2 - V(\phi, X) \ ,
\end{align}
where the potential $V(\phi,X)$ is given in Eq.~\eqref{eq:model}. It is convenient to write these in terms of the natural variables \eqref{eq:newvars2}-\eqref{eq:newvars1} as follows,
\begin{align} 
\rho  &\equiv \frac{\omega_*^2 \phi_*^2}{a^{3/2}}\left[ \sum_f^{\varphi,\chi}(E^f_{\rm k} +E^f_{\rm g} + E^f_{\rm p}) + E_{\rm i}\right], \\
    p&\equiv \frac{\omega_*^2 \phi_*^2}{a^{3/2}}\left[\sum_f^{\varphi,\chi}( E^f_{\rm k}-\frac{1}{3}E^f_{\rm g} -E^f_{\rm p}) - E_{\rm i} \right], \label{eq:EnPresExp} 
\end{align}
where $E_{\rm k}^{f}$, $E_{\rm g}^{f}$ and $E_{\rm p}^{f}$ (with $f=\varphi,\chi$) are the \textit{kinetic}, \textit{gradient} and \textit{potential} contributions of each field, and $E_{\rm i}$ the \textit{interaction} contribution, defined as follows,
\begin{align}
    E_{\rm k}^f &\equiv   \frac{1}{2a^2}\left( f' - \frac{3a'}{2 a} \right)^2  \ ,  \hspace{0.2cm}  &E_{\rm g}^f &\equiv  \frac{1}{2a^2}|\vec{\nabla}_{\vec y} f|^2 \ , \label{eq:Ecomp0} \\
   E_{\rm p}^{\varphi} &\equiv   \frac{ a^3}{\omega_*^2 \phi_*^2}V_{\rm inf} (\varphi) \ ,    \hspace{0.2cm}   &E_{\rm p}^{\chi} &\equiv \frac{1}{4} \tilde{q}^{(\lambda)} \chi^4  \ , \\
    E_{\rm i}&\equiv  \frac{1}{2} \tilde{q}^{(h)} \varphi \chi^2  \ .  \label{eq:Ecomp}
\end{align}
We can hence define the total energy density of the inflaton and daughter field as follows,
\begin{align}
E^\varphi&=E_{\rm k}^{\varphi} + E_{\rm g}^{\varphi} +  E_{\rm p}^{\varphi} + E_{\rm i}  \ ,\\
E^\chi&=E_{\rm k}^{\chi} + E_{\rm g}^{\chi} + E_{\rm p}^{\chi} + E_{\rm i} \label{eq:linEchi} \ .
\end{align}
It is also convenient to define, for each energy contribution $E_{\alpha}$, its corresponding \textit{energy ratio} as $\varepsilon_{\alpha} \equiv E_{\alpha} / \sum_{\beta} E_{\beta}$. By construction, all energy ratios sum to one, $\sum_{\beta} \varepsilon_{\beta} = 1$.

 \subsection{Comparison of simulations in 2+1 and 3+1 dimensions}\label{App:3D}

In the left panel of Fig.~\ref{fig:Specs2D3D} we compare the evolution of the daughter field spectrum obtained from lattice simulations in 3+1 (dashed lines) and 2+1 dimensions (solid lines), for the choice of parameters $q_*^{(h)}=50$ and $q_*^{(\lambda)}=2500$, obtained with a lattice of $N=256$ points per dimension and infrared momentum cutoff $k_{\rm ir}=2.9$. The spectral evolution is very similar during both the early resonance regime and the late non-linear stage. In particular, the $2+1$-dimensional simulations reproduce well the position and amplitude of the main spectral peak at late times appearing in the $3+1$-dimensional ones. On the other hand, in the right panel of Fig.~\ref{fig:Specs2D3D} we compare the evolutions of the equation of state $w$. They also coincide very well, and in particular, both simulations reproduce the early jump of the equation of state around the backreaction time $u\sim10$, as well as the later relaxation towards $\bar{w} \rightarrow 0$.

\subsection{Coupling constants values} \label{sec:CoupConsts}

The relations between the coupling parameters $q_*^{(h)}$ and $q_*^{(\lambda)}$ and the coupling constants $h$ and $\lambda$ are given by Eqs.~\eqref{eq:respar} and \eqref{eq:resself}. For the $\alpha$-attractor T-model \eqref{eq:alphaattractor}, which we have investigated with $M=5\mpl$ and $N_k=55$, they are related as follows, 
\begin{align}
    h & \equiv \left( \frac{\omega_*^2}{\phi_*}\right) q_*^{(h)} = 3.0 \cdot 10^{-11} \, q_*^{(h)} \, \mpl , \label{eq:transA} \\
    \lambda &\equiv \left( \frac{\omega_*^2}{\phi_*^2} \right)  q_*^{(\lambda)} = 2.2 \cdot10^{-11}\,q_*^{(\lambda)} \label{eq:transB} \ ,
\end{align}
where we have obtained $\omega_*^2\equiv \Lambda^4/M^2$ with $\Lambda^4(M,N_k)=1.01 \cdot 10^{-9}\mpl^4$ from Eq.~\eqref{eq:Lambda4} and $\phi_*=\phi_e$ is given by Eq.~\eqref{eq:phiend}.

\subsection{GW suppression factor}\label{App:GWDilution}

Here we explain how to redshift the GW spectrum from preheating from the time the gravitational wave production ends $u_{\rm f}$ until the present time $u_0$. Let us consider some physical frequency today,
\be
f_{\rm gw}=\frac{k}{2\pi a_0} \ . \label{Eq:fgw_App}
\ee
We can express $a_0$ in the following way
\be
\frac{1}{a_0}=\frac{1}{a_{\rm f}H_{\rm f}}\frac{a_{\rm f}}{a_{\rm e}}\frac{a_{\rm e}}{a_{\rm rd}}\frac{a_{\rm rd}}{a_0}H_{\rm f}\ ,  \label{Eq:a0_App}
\ee
where $a_{\rm e}\equiv a(u_{\rm e})$ denotes the scale factor at the end of inflation, $a_{\rm f} \equiv a(u_{\rm f} )$ at the final time of GW production, $a_{\rm rd}\equiv a(u_{\rm rd})$ at the onset of radiation domination, and $a_0$ today. This expression propagates the frequency from the final time of GW production $u_{\rm f}$ back to the end of inflation $u_{\rm e}$, and then forward to the onset of radiation domination $u_{\rm rd}$. By assuming that the universe thermalizes at the onset of radiation domination (otherwise we would need to incorporate another small correction factor that accounts for the change of degrees of freedom between $u_{\rm rd}$ and the time of thermalization $u_{\rm th}$), we can express the last ratio in terms of the corresponding energy densities and ratios of degrees of freedom
\be
\frac{1}{a_0}=\frac{1}{a_{\rm f}H_{\rm f}}\frac{a_{\rm f}}{a_{\rm e}}\frac{a_{\rm e}}{a_{\rm rd}}{\mathcal{G}}^{1/4}\rho_{\rm rd}^{-1/4}\rho_0^{1/4}H_{\rm f}\ ,   \label{Eq:a01_App}
\ee
where ${\mathcal{G}}=(g_{\rm rd}/g_0)(g_{s,0}/g_{0,\rm rd})^{4/3}$, with $g_{\rm rd}$ and $g_{0}$ being the relativistic degrees of freedom at radiation domination and today respectively. The energy density at the moment of radiation domination $\rho_{\rm rd}$ can be expressed in terms of the energy density at the final time of GW production $\rho_{\rm f}$ in the following way,
\be
\rho_{\rm rd}=e^{-3(1-\bar{w}_{\rm rd})N_{\rm rd}+3(1-\bar{w}_{\rm i})N_{\rm i}}\rho_{\rm f}\, ,  \label{Eq:rhof_App}
\ee
where
\bea
\bar{w}_{\rm rd} &=& \frac{1}{N_{\rm rd}} \int_0^{N_{\rm rd}} d N' w (N')\ , \\
\bar{w}_{\rm i} &=&\frac{1}{N_{\rm i}} \int_0^{N_{\rm i}} d N' w (N')\ ,
\eea
with $N_{\rm r} \equiv \ln ( a_{\rm rd}/a_{\rm e})$ and $N_{\rm i}\equiv \ln ( a_{\rm f}/a_{\rm e})$. Again, in this expression the value of $\rho_{\rm f}$ is first propagated back to the end of inflation and then forward to the onset of radiation domination. Inserting Eq.~(\ref{Eq:rhof_App}) into (\ref{Eq:a01_App}) yields,
\be
\frac{1}{a_0}=\frac{1}{a_{\rm f}H_{\rm f}}\epsilon_{\rm i}^{-1/4}\epsilon_{\rm rd}^{1/4}{\mathcal{G}}^{1/4}\rho_{\rm f}^{-1/4}\rho_0^{1/4}H_{\rm f}\ ,
\ee
with
\be
\epsilon_{\rm i}=\left(\frac{a_{\rm e}}{a_{\rm f}}\right)^{1-3\bar{w}_{\rm i}}\ , \,\,\,\, \epsilon_{\rm rd}=\left(\frac{a_{\rm e}}{a_{\rm rd}}\right)^{1-3\bar{w}_{\rm rd}}\ .
\ee
Inserting this quantity back into Eq.~\eqref{Eq:fgw_App} we obtain,
\begin{align}
f_{\rm gw}&=\frac{{\mathcal{G}}^{1/4}\rho_0^{1/4}}{3^{1/4}2\pi}\epsilon_{\rm i}^{-1/4}\epsilon_{\rm rd}^{1/4}\frac{k}{a_{\rm f}H_{\rm f}}\left(\frac{H_{\rm f}}{\mpl}\right)^{1/2}\\
&=4\cdot10^{10}\epsilon_{\rm i}^{-1/4}\epsilon_{\rm rd}^{1/4}\frac{k}{a_{\rm f}H_{\rm f}}\left(\frac{H_{\rm f}}{\mpl}\right)^{1/2} {\rm Hz} \ ,
\end{align}
where we have used today's radiation density $\rho_0=\rho_{0,\rm rad}\approx3.4\cdot10^{-51}\rm GeV^4$ and $g_{\rm rd}=g_{s, \rm rd}=10^2$.

\bibliography{References.bib,extra.bib}
  \bibliographystyle{utphys}

\end{document}